\documentclass[twocolumn,twocolappendix]{aastex631}

\usepackage{graphicx}	
\usepackage{amsmath}	
\usepackage{subfigure}
\usepackage{natbib}
\usepackage{color}
\usepackage{tabularx}
\usepackage{longtable}
\usepackage{bm}
\usepackage{threeparttable}
\usepackage{enumerate}
\usepackage{chemfig}
\usepackage[version=4,arrows=pgf-filled,
textfontname=sffamily,
mathfontname=mathsf]{mhchem}
\usepackage[normalem]{ulem}
\usepackage{verbatim}
\usepackage[T1]{fontenc}

\newcommand{\code}[1]{{\texttt{#1}}}

\newcommand{\gaia}{{\it Gaia}}

\newcommand{\tar}{{HAT-P-70}}
\newcommand{\angstrom}{\text{\normalfont\AA}}

\begin{document}

\title{Revisiting the atmosphere of HAT-P-70b with CARMENES high-resolution transmission spectroscopy}

\correspondingauthor{Tianjun Gan, Jaume Orell-Miquel}
\email{tianjungan@gmail.com, jaume.miquel@austin.utexas.edu}

\author[0000-0002-4503-9705]{Tianjun~Gan}
\affil{Department of Astronomy, Westlake University, Hangzhou 310030, Zhejiang Province, China}

\author[0000-0003-2066-8959]{Jaume Orell-Miquel}
\affil{Department of Astronomy, University of Texas at Austin, 2515 Speedway, Austin, TX 78712, USA}

\author[0000-0001-9585-9034]{Fei Yan}
\affil{ Department of Astronomy, University of Science and Technology of China, Hefei 230026, China}

\author[0000-0001-8419-8760]{Lisa Nortmann}
\affil{Institut f\"ur Astrophysik und Geophysik, Georg-August-Universit\"at G\"ottingen, Friedrich-Hund-Platz 1, 37077 G\"ottingen, Germany}

\author[0000-0002-1600-7835]{Jorge Sanz-Forcada}
\affil{Centro de Astrobiolog\'ia (CSIC-INTA), Camino bajo del castillo s/n, ESAC Campus, 28692 Villanueva de la Cañada, Madrid, Spain}

\author[0000-0003-0987-1593]{Enric Pall{\'e}}
\affil{Instituto de Astrof\'isica de Canarias (IAC), V\'ia L\'actea s/n, E-38205 La Laguna, Tenerife, Spain}
\affil{Departamento de  Astrof\'isica, Universidad de La Laguna (ULL), E-38206 La Laguna, Tenerife, Spain}

\author[0000-0001-8317-2788]{Shude Mao} 
\affil{Department of Astronomy, Westlake University, Hangzhou 310030, Zhejiang Province, China}


\author[0000-0002-8388-6040]{Pedro~J.~Amado}
\affil{Instituto de Astrof\'isica de Andaluc\'ia (IAA-CSIC), Glorieta de la Astronom\'ia s/n, 18008 Granada, Spain}

\author[0000-0002-7349-1387]{Jos\'e~A.~Caballero}
\affil{Centro de Astrobiolog\'ia (CSIC-INTA), Camino bajo del castillo s/n, ESAC Campus, 28692 Villanueva de la Cañada, Madrid, Spain}

\author[0000-0002-7671-2317]{Stefan Cikota}
\affil{Centro Astron\'omico Hispano en Andaluc\'ia, Observatorio de Calar Alto, Sierra de los Filabres, 04550 G\'ergal, Almer\'ia, Spain}

\author[0000-0002-6187-8154]{David Cont}
\affil{Universit\"ats-Sternwarte, Ludwig-Maximilians-Universit\"at
M\"unchen, Scheinerstrasse 1, 81679 M\"unchen, Germany}
\affil{Exzellenzcluster Origins, Boltzmannstrasse 2, 85748 Garching bei M\"unchen, Germany}

\author[0000-0002-3404-8358]{Artie P. Hatzes}
\affil{Th\"uringer Landessternwarte Tautenburg, Sternwarte 5, 07778 Tautenburg, Germany}

\author[0000-0002-1493-300X]{Thomas Henning}
\affil{Max-Planck-Institut f\"ur Astronomie (MPIA), K\"onigstuhl 17, 69117 Heidelberg, Germany}

\author[0000-0001-6082-1175]{Fabio Lesjak}
\affil{Leibniz Institute for Astrophysics Potsdam (AIP), An der Sternwarte 16, 14482 Potsdam, Germany}

\author[0000-0003-2941-7734]{Manuel L\'opez-Puertas}
\affil{Instituto de Astrof\'isica de Andaluc\'ia (IAA-CSIC), Glorieta de la Astronom\'ia s/n, 18008 Granada, Spain}

\author[0000-0002-7779-238X]{David Montes}
\affil{Departamento de F\'isica de la Tierra y Astrof\'isica, Facultad de Ciencias F\'isicas, e IPARCOS-UCM (Instituto de F\'isica de Part\'iculas y del Cosmos de la UCM), Universidad Complutense de Madrid, 28040 Madrid, Spain}

\author[0000-0003-0061-518X]{Juan Carlos Morales}
\affil{Institut de Ci\'encies de l'Espai (CSIC-IEEC), Campus UAB, c/de Can Magrans s/n, 08193 Bellaterra, Barcelona, Spain}
\affil{Institut d'Estudis Espacials de Catalunya (IEEC), 08034 Barcelona, Spain}

\author[0000-0001-9204-8498]{Alberto Pel\'aez-Torres}
\affil{Instituto de Astrof\'isica de Andaluc\'ia (IAA-CSIC), Glorieta de la Astronom\'ia s/n, 18008 Granada, Spain}

\author[0000-0002-3302-1962]{Andreas Quirrenbach}
\affil{Landessternwarte, Zentrum f\"ur Astronomie der Universit\"at Heidelberg, K\"onigstuhl 12, 69117 Heidelberg, Germany}

\author[0000-0003-1242-5922]{Ansgar Reiners}
\affil{Institut f\"ur Astrophysik und Geophysik, Georg-August-Universit\"at G\"ottingen, Friedrich-Hund-Platz 1, 37077 G\"ottingen, Germany}

\author[0000-0002-6689-0312]{Ignasi Ribas}
\affil{Institut de Ci\'encies de l'Espai (CSIC-IEEC), Campus UAB, c/de Can Magrans s/n, 08193 Bellaterra, Barcelona, Spain}
\affil{Institut d'Estudis Espacials de Catalunya (IEEC), 08034 Barcelona, Spain}

\author[0000-0002-1624-0389]{Andreas Schweitzer}
\affil{Hamburger Sternwarte, Gojenbergsweg 112, 21029 Hamburg, Germany}

\begin{abstract}

Owing to hot and inflated envelopes that facilitate atmospheric studies, ultra-hot Jupiters (UHJs) have attracted much attention. Significant progress has been achieved, from enlarging the sample size to broadening the studies to encompass diverse stellar types and ages. Here, we present a transmission spectroscopy study of \tar b, an UHJ orbiting a young A-type star, through high-resolution observations with CARMENES at the 3.5\,m Calar Alto telescope. By using the line-by-line technique, we confirm the previous detections of H$\alpha$, Na\,{\sc i}, and Ca\,{\sc ii}, report a new tentative detection of K\,{\sc i}, and impose an upper limit on the He triplet absorption. Through cross-correlation analysis, we identify the Ca\,{\sc ii} and Fe\,{\sc i} absorptions, both blue-shifted by approximately $\rm 5\,km\ s^{-1}$, indicating a day-to-night side atmospheric wind. Additionally, we find a new tentative detection of K\,{\sc i}. We do not see any significant atmospheric molecular signal in the near-infrared data. Putting \tar b in the context of UHJs from the literature, it turns out that (1) H$\alpha$ absorption is more common on gas giants orbiting stars younger than 1\,Gyr, with a relative detection probability of $P_{\rm Age<1\,Gyr}({\rm H}\alpha)/P_{\rm Age\geq1\,Gyr}({\rm H}\alpha)\sim 3$; (2) any UHJ is likely to exhibit Fe\,{\sc i} absorption if it has Ca\,{\sc ii}. 




\end{abstract}



\section{Introduction} \label{section_introduction}


Among all planetary classes, ultra-hot Jupiters (UHJs)$-$defined as gas giants with equilibrium temperatures $T_{\rm eq}$ above about 2200~K \citep{Parmentier2018,Stangret2022,Maiz2024}$-$serve as benchmark systems for atmospheric characterization, which benefit from their large scale height due to the proximity to their hot host stars. The majority of UHJs are supposed to be largely cloud-free on the dayside hemisphere, since the extreme insolation inhibits condensation processes \citep{Helling2019,Gao2020}, making them ideal targets for atmospheric composition studies. Meanwhile, the atomic and molecular species undergo thermal dissociation and ionization \citep{Arcangeli2018,Lothringer2018} such that metals, particularly refractory elements, either neutral or ionized, exist in the gaseous state and thus could be observed. Detections of these species, and the measurement of their abundances through atmospheric retrievals \citep{Madhusudhan2009,Brogi2019,Gibson2020}, offer a unique opportunity to test theoretical predictions, probing planet formation histories \citep[e.g.,][]{Oberg2011,Mordasini2016,Turrini2021,Molliere2022,Chachan2023}, as well as subsequent migration processes \citep[e.g.,][]{Madhusudhan2014,Lothringer2021,Schneider2021}. Given their short orbital periods, UHJs are likely tidally locked. Such a special orbital configuration would result in a significant day- and night-side temperature contrast, which drives atmospheric heating gradients and wind flows \citep{Kempton2012,Bell2018}, a possible explanation for the observed chemical diurnal inhomogeneities in some planetary atmospheres \citep{Ehrenreich2020,Cont2021,Cont2024,Prinoth2022,Maguire2024}. 

One route to characterize planetary atmospheres is through the transmission spectroscopy methodology \citep{Seager2000}. The light from the host star is blocked by the planet atmosphere, imprinting wavelength-dependent signatures on the observed spectrum. Therefore, it contains information on the morning/evening terminator atmospheres. In particular, the high-resolution transmission spectroscopy \citep[see the review by][and references therein]{Snellen2025} can trace spectral absorption of individual lines, or map the signals from multiple lines onto a single detection peak through cross-correlation \citep{Snellen2010}. Over the past two decades, numerous UHJs' atmospheres have been investigated with this approach (see Table~1 in \citet{Cont2024}), revealing a wealth of atomic species such as Ca\,{\sc i}, Cr\,{\sc i}, Fe\,{\sc i}, Na\,{\sc i}, K\,{\sc i}, Ti\,{\sc i}, and V\,{\sc i} \citep[e.g.,][]{Hoeijmakers2019,Yan2019,Chen2020,Casasayas-Barris2022,Langeveld2022,Zhang2022,Borsato2023,Stangret2024,Cont2025,Vaulato2025}, as well as molecules such as $\rm CO$, $\rm H_2O$, $\rm OH$, $\rm TiO$, and $\rm VO$ \citep[e.g.,][]{Landman2021,Sedaghati2021,Prinoth2022,Pelletier2023,Gandhi2024,Hood2024,Maguire2024,Mansfield2024,Mraz2024,Simonnin2025}.

Optical high-resolution transmission spectroscopy is primarily sensitive to refractory metal elements as they possess substantial absorption lines in the blue wavelength band \citep[i.e., 3500--5500~\AA;][]{Kurucz2018}. The absorption of molecules that consist of volatile elements such as carbon, oxygen, and sulfur, however, predominantly appear in the near-infrared regime \citep{Tennyson2016}. Hence, the combination of optical and near-infrared spectroscopic observations enables a thorough mapping of the atmosphere, for example, determining the refractory-to-volatile ratio, which will in turn shed light on the birthplace of the planet \citep{Lothringer2021}. In addition, the H$\alpha$ $\lambda$6565\,{\AA} and He~{\sc i} $\lambda$10\,833\,{\AA} lines
(all wavelengths throughout the paper are defined in vacuum) are key diagnostics of atmospheric escape \citep{Yan2018_KELT-9b, Nortmann_WASP-69_He, Kirk2020,Lampon_2021_regimenes,Gully2024,Saidel2025}. Both lines have different observational advantages \citep{Seager2000, Fuhrmeister2020_He_variability, Orell-Miquel2022}, so simultaneously observing H$\alpha$ and He\,{\sc i} triplet lines offers a more comprehensive view of atmospheric escape \citep{Orell-Miquel2024}. Hydrodynamic modeling of the H$\alpha$ and He\,{\sc i} line profiles can retrieve the mass loss rate \citep{Lampon_2021_regimenes, HAT-P-32b_Zhang2023}, which is critical for understanding the evolutionary path and fate of the exoplanet.

In this work, we investigate the atmosphere of \tar b (see Table~\ref{param}), an UHJ orbiting a young (0.6\,Gyr) A-type star every 2.7 days \citep{Zhou2019}, with CARMENES high-resolution transmission spectroscopy. An initial study by \cite{BelloArufe2022} using HARPS-N ($R \sim 115\,000$, 3830--6930~\AA) detected nine atomic species in total including Ca\,{\sc ii}, Cr\,{\sc i}, Cr\,{\sc ii}, Fe\,{\sc i}, Fe\,{\sc ii}, H\,{\sc i}, Mg\,{\sc i}, Na\,{\sc i}, and V\,{\sc i}. Absorption from the Ca\,{\sc ii} infrared triplet (IRT) was also recently reported by \cite{Langeveld2025} using the Gemini High-resolution Optical SpecTrograph (GHOST; $R \sim 76\,000$, 3830--10\,000~\AA). With CARMENES observations, we revisit the optical atmospheric signatures of \tar b and extend the transmission spectroscopy to the optical red and near-infrared bands. 

The paper is structured as follows. In Section~\ref{section_observations}, we describe the observations and data reduction. Section~\ref{section_analysis} details our analysis including single-line and cross-correlation methods. We summarize our results in Section~\ref{section_results}. We discuss our findings in Section~\ref{section_discussions} before we conclude in Section~\ref{section_conclusions}.

\section{Observations}\label{section_observations}

We observed one transit of \tar b on UT 2023 November 26 using the CARMENES spectrograph \citep{Quirrenbach2014,Quirrenbach2018}. The observation was carried out as part of the CARMENES Legacy program (23B-3.5-103). CARMENES is a high-resolution echelle spectrograph installed at the 3.5\,m telscope at the Observatorio de Calar Alto in Almer\'ia, Spain. The observation was performed simultaneously in two channels (VIS and NIR), covering the wavelength ranges 0.52–-0.96\,$\mu$m and 0.96–-1.71\,$\mu$m with spectral resolutions of 94\,600 and 80\,400, respectively. 

\begin{figure}[]
    \centering
    \includegraphics[width=\linewidth]{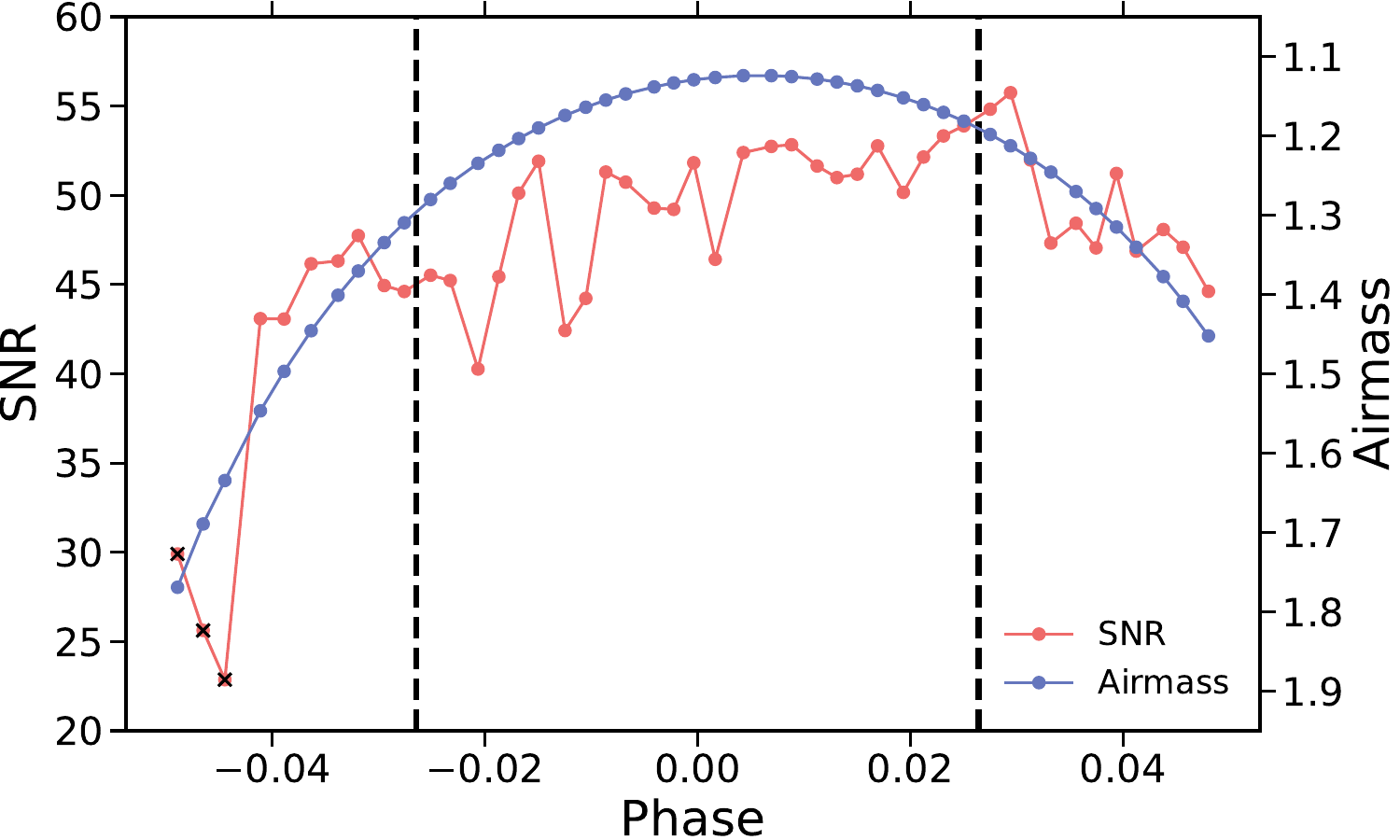}
    \caption{The S/N of the spectra (red) near the Na\,{\sc i} $\lambda\lambda$5891,5898\,$\angstrom$ doublet lines 
    and the airmass (blue) as a function of planet phase. Two vertical black dashed lines mark the first and fourth contacts of the transit event. The first three spectra (black crosses) are discarded due to their low S/N.}
    \label{fig:SNR_airmass}
\end{figure}

Over the duration of 6.4\,h of the observation, a total of 46 spectra were collected in each channel with a fixed exposure time of 400\,s, covering the 3.5\,h-long transit as well as 1.5\,h before the ingress and 1.4\,h after the egress. During the observations, the airmass varied between 1.12 and 1.77. Figure~\ref{fig:SNR_airmass} shows the signal-to-noise ratio (S/N) per pixel near the Na\,{\sc i} doublet lines at 5891~\AA\ and 5898~\AA\ along with the airmass trend. The S/N of the first three spectra is much lower than that of the others due to high airmass and poor seeing conditions. Consequently, we excluded these spectra in the following analysis. The raw CARMENES spectra were reduced with the \code{caracal} pipeline \citep{Caballero2016} and the calibrated 1D spectra were obtained via the flat-optimized extraction algorithm \citep{Zechmeister2014}. 

\begin{table}[h]
    \centering
    \caption{Stellar and planetary parameters of the \tar b system.}
    \begin{tabular}{lcc}
        \hline\hline
        Parameter       &Value       &Ref. \\\hline
        \it{Main identifiers} \\
        HD  &287325 &[1]\\
        \gaia\ ID            &$3291455819447952768$ &[2]\\
        \it{Equatorial coordinates} \\
        $\alpha_{\rm J2000}$    &\ \ 04:58:12.56 &[2]\\
         $\delta_{\rm J2000}$    &+09:59:52.7   &[2]\\
        \it{Stellar properties}                    \\
        $d$ [pc] &$318.2\pm2.4$ &[2]\\
        $G$ [mag] &$9.458\pm0.003$ &[2] \\
        $J$ [mag] &$9.068\pm0.022$ &[3]\\
        $H$ [mag] &$9.023\pm0.029$ &[3]\\
        $K_s$ [mag] &$8.963\pm0.024$ &[3]\\
         $M_\ast$ [$\rm M_\odot$] &$1.890^{+0.010}_{-0.013}$ &[4]\\
         $R_\ast$ [$\rm R_\odot$] &$1.858^{+0.119}_{-0.091}$ &[4]\\
         $\log g_\star$ [cgs] &$4.181^{+0.055}_{-0.063}$ &[4]\\
         $T_{\rm eff}$ [K] &$8450^{+540}_{-690}$ &[4]\\
         $\rm [Fe/H]$ [dex] &$-0.059^{+0.075}_{-0.088}$ &[4]\\
         $v\sin i$ [$\rm km\ s^{-1}$] &$99.85^{+0.64}_{-0.61}$ &[4]\\
         $\gamma$ [$\rm km\ s^{-1}$] &$+25.26\pm0.11$ &[4]\\
         Age [Gyr] &$0.60^{+0.38}_{-0.20}$ &[4]\\
        \it{Planetary parameters} \\
        $P$ [d] &$2.74432452^{+0.00000079}_{-0.00000068}$ &[4]\\
        $a$ [au] &$0.04739^{+0.00031}_{-0.00106}$ &[4]\\
        $T_{\rm 14}$$^{[5]}$ [d]  &$0.1450_{-0.0020}^{+0.0028}$ &[4]\\
        $K$ [$\rm m\ s^{-1}$] &$<649\ (3\sigma)$ &[4]\\
        $i$ [deg] &$83.50^{+0.91}_{-1.42}$ &[4]\\
        $R_p$ [$\rm R_{Jup}$] &$1.87^{+0.15}_{-0.10}$ &[4]\\
        $M_p$ [$\rm M_{Jup}$] &$<6.78\ (3\sigma)$ &[4]\\
        $\rho_p$ [$\rm g\ cm^{-3}$] &$<1.54\ (3\sigma)$ &[4]\\
        $\log g_p$ [cgs] &$<3.73\ (3\sigma)$ &[4]\\
        $T_{\rm eq}$$^{[6]}$ [K] &$2562^{+43}_{-52}$ &[4]\\
        $K_p$ [$\rm km\ s^{-1}$] &$186.7\pm4.2$ &[4]\\
         \hline
    \end{tabular}
    \begin{tablenotes}
    \item[1]  [1]~\cite{Cannon1918}; [2]~\cite{Gaia2023}; [3]~\cite{skrutskie2006}; [4]~\cite{Zhou2019}; [5]~$T_{\rm 14}$ represents the total transit duration time between first and last contact; [6]~$T_{\rm eq}$ was calculated assuming zero albedo and full heat redistribution.
    \end{tablenotes}
    \label{param}
\end{table}

\section{Analysis}\label{section_analysis}

In this section, we utilized single-line and cross-correlation methods to search for atomic and molecular species in the atmosphere of \tar b. To summarize, we looked for single-line absorption of H$\alpha$, the He\,{\sc i} NIR triplet, the Ca\,{\sc ii}\,IRT, the Na\,{\sc i} doublet, Li\,{\sc i} $\lambda$6710\,\AA, and K\,{\sc i} $\lambda$7701\,\AA; and we investigated 14 neutral/ionized species including Ca\,{\sc i}, Ca\,{\sc ii}, Cr\,{\sc i}, Fe\,{\sc i}, Fe\,{\sc ii}, K\,{\sc i}, Li\,{\sc i}, Mg\,{\sc i}, Na\,{\sc i}, Si\,{\sc i}, Ti\,{\sc i}, V\,{\sc i}, V\,{\sc ii}, Y\,{\sc i} as well as 11 molecules: $\rm CH_{4}$, $\rm CO$, $\rm CO_2$, $\rm FeH$, $\rm H_{2}O$, $\rm H_{2}S$, $\rm HCN$, $\rm NH_3$, $\rm OH$, $\rm TiO$, and $\rm VO$ through cross-correlation. The selected species above have considerable number of lines or strong lines within the wavelength range of CARMENES, leading to a higher detection sensitivity compared with others. 

\subsection{Rossiter–McLaughlin and Center-to-Limb Variation Effect Correction}

The rapid rotation of the host star and limb darkening introduce potential contamination from the Rossiter-McLaughlin \citep[RM;][]{Rossiter1924,McLaughlin1924} and center-to-limb variation \citep[CLV;][]{Czesla2015,Yan2017,Casasayas-Barris2018,Chen2020,Bello-Arufe2023,Fossati2023,Maguire2023,Biassoni2024,Sicilia2025} effects, which could distort stellar lines and alter the strength of planetary signals or even mimic planetary signals. Therefore, we modeled the RM+CLV effects for the HAT-P-70b transit following the method described in \cite{Yan2018_KELT-9b}. We first simulated the stellar spectrum at different limb angles using the Spectroscopy Made Easy tool \citep{Piskunov2016} with the VALD line list \citep{Ryabchikova2015} and the ATLAS12 model \citep{Kurucz1996}. We computed the disk-averaged spectrum (4000--10\,000~\AA) during transit, assuming a stellar rotation velocity ($v\sin i$) of 98.6\ ${\rm km\ s^{-1}}$ and a spin-orbit angle of 107.9\,deg \citep{BelloArufe2022}. The model was then normalized by the out-of-transit model spectrum, which is not impacted by these two effects. Figure~\ref{fig:RM_CLV_model} shows the RM+CLV model around the Na\,{\sc i} doublet. The overlap between the model and the planetary orbital motion at the beginning of the transit event highlights the necessity of the RM and CLV effect correction. We applied the normalized RM+CLV model in the following analysis.

\subsection{Single-Line Analysis}\label{single_line_analysis}

We computed the single-line transmission spectrum (TS) from the CARMENES VIS and NIR spectra.
We mainly followed the methodology described by \cite{Orell-Miquel2022, Orell-Miquel2023, Orell-Miquel2024} based on the well-established procedure presented by \citet{Wyttenbach2015}.
In each case, we worked with a relatively short spectral range centered on the line of interest.
The He\,{\sc i} triplet is the only inspected line with telluric emission lines of OH in its surroundings. For each pair of target (fiber A) and sky (fiber B) spectra, we divided the science spectrum by its particular OH emission model, generated from fitting simultaneously the three main OH peaks with three independent Gaussian profiles. We accounted for the different efficiency between the two fibers.
For all the inspected lines, we used a telluric model from \texttt{molecfit} \citep{molecfit_1}, which accounts for the instrumental profile and spectral resolution, to empirically fit and correct the surrounding telluric absorption \citep{Nuria2017, Yan2019}.

\begin{figure}[]
    \centering
    \includegraphics[width=0.49\textwidth]{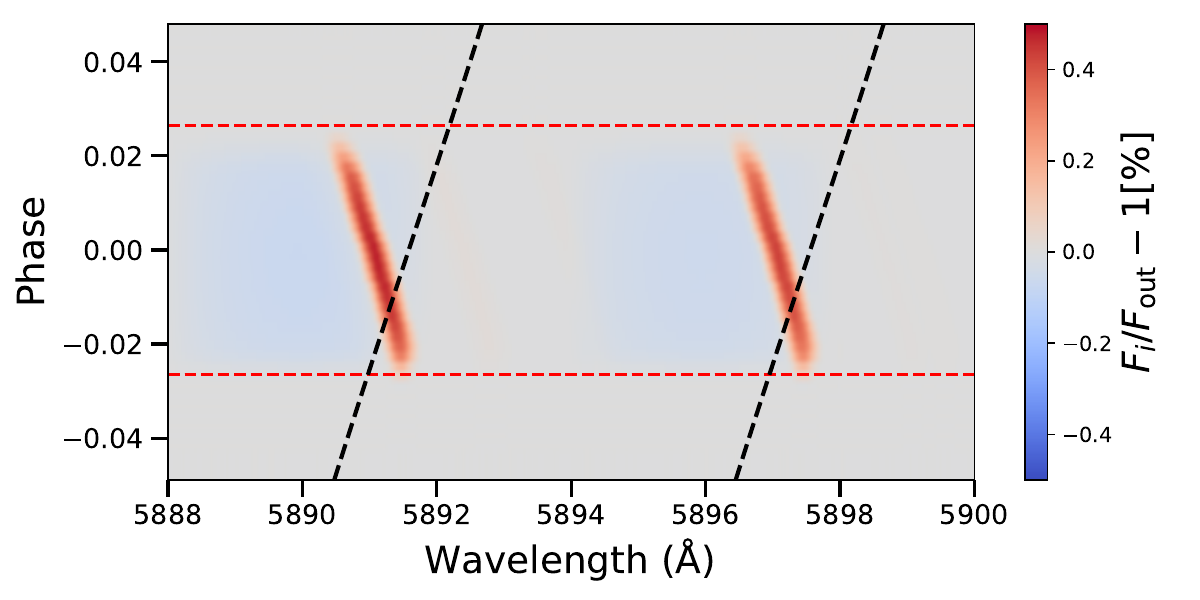}
    \caption{The model of the RM and CLV 
    (see Section~\ref{single_line_analysis}) around the Na\,{\sc i} doublet, normalized by the out-of-transit model spectrum, in the stellar rest frame. The horizontal red dashed lines mark the first and fourth contacts of the transit event. The black tilted dashed line represents the trace of the planetary orbital motion.}
    \label{fig:RM_CLV_model}
\end{figure}

Once corrected from the telluric contributions, we normalized each spectrum by its own continuum. Then, we shifted the spectra into the stellar rest frame and created a high S/N stellar spectrum by averaging all the out-of-transit spectra. We divided all spectra by this stellar spectrum to remove the stellar contribution. At this step, we also corrected the RM+CLV effect by interpolating the model into the dataset grid and dividing the observations by the interpolated model. The RM+CLV effect on the TS of the inspected lines is of the same order of the TS uncertainties ($\sim$0.4\,\%, see Figure\,\ref{fig:RM_CLV_model}). Next, we shifted the spectra onto the planet rest frame. We computed the TS as the average using the inverse of the squared errors as weights.

Finally, we used a Gaussian profile to fit any planetary signal in the TS and measure the absorption depth (and equivalent width, EW), the width $\sigma$ (and full width at half maximum, FWHM), and velocity shift $\Delta v$ of the signal. We employed the MultiNest algorithm (\citealp{MultiNest}) via its {\tt python} implementation \texttt{PyMultinest} \citep{PyMultiNest}.

\subsection{Cross-Correlation Analysis}


We next performed a cross-correlation analysis \citep{Snellen2010} following the procedures described in \cite{Stangret2020,Stangret2022}. We outline the data pre-processing, template model generation and the details of cross-correlation below. 

\subsubsection{Spectra Pre-Processing}\label{spec_pre_processing}

Since a few hundred pixels of the first ten orders of the VIS-band spectra lack flux information, we fixed missing values of the initial 700 pixels of them to the median flux. We flagged all other pixels of the spectra with flux values saved as NaN (Not a Number), and replaced their fluxes and uncertainties with the interpolated values using a cubic spline model. We next excluded the outliers, for example due to cosmic rays, in the data. To do this, we explored the time evolution of each pixel of each order. We fitted a cubic spline function, and replaced the flux with the best-fit result if it is discrepant with the model by $5\sigma$. Normalization was then performed order by order. Basically, we split each order into several chunks with a binning size of 50 pixels and obtained the averaged continuum by computing the median wavelength and flux of the top 10 flux values within each chunk \citep{Passegger2016,Hoeijmakers2020}. We interpolated the continuum to the observational wavelength stamps and normalized the spectrum through dividing the data by the interpolated continuum to calibrate the baselines to a similar flux level. Finally, we mitigated sky emission lines with flux exceeding 10 percent of the continuum level and strong absorption lines of tellurics with normalized flux below 0.1 by artificially masking them. To correct the RM and CLV effects, we interpolated the RM+CLV model (see Section~\ref{single_line_analysis}) to the wavelength and orbital phase of the observations, and divided the spectrum matrix by the RM+CLV model order by order. 

\begin{figure}[]
    \centering
    \includegraphics[width=0.5\textwidth]{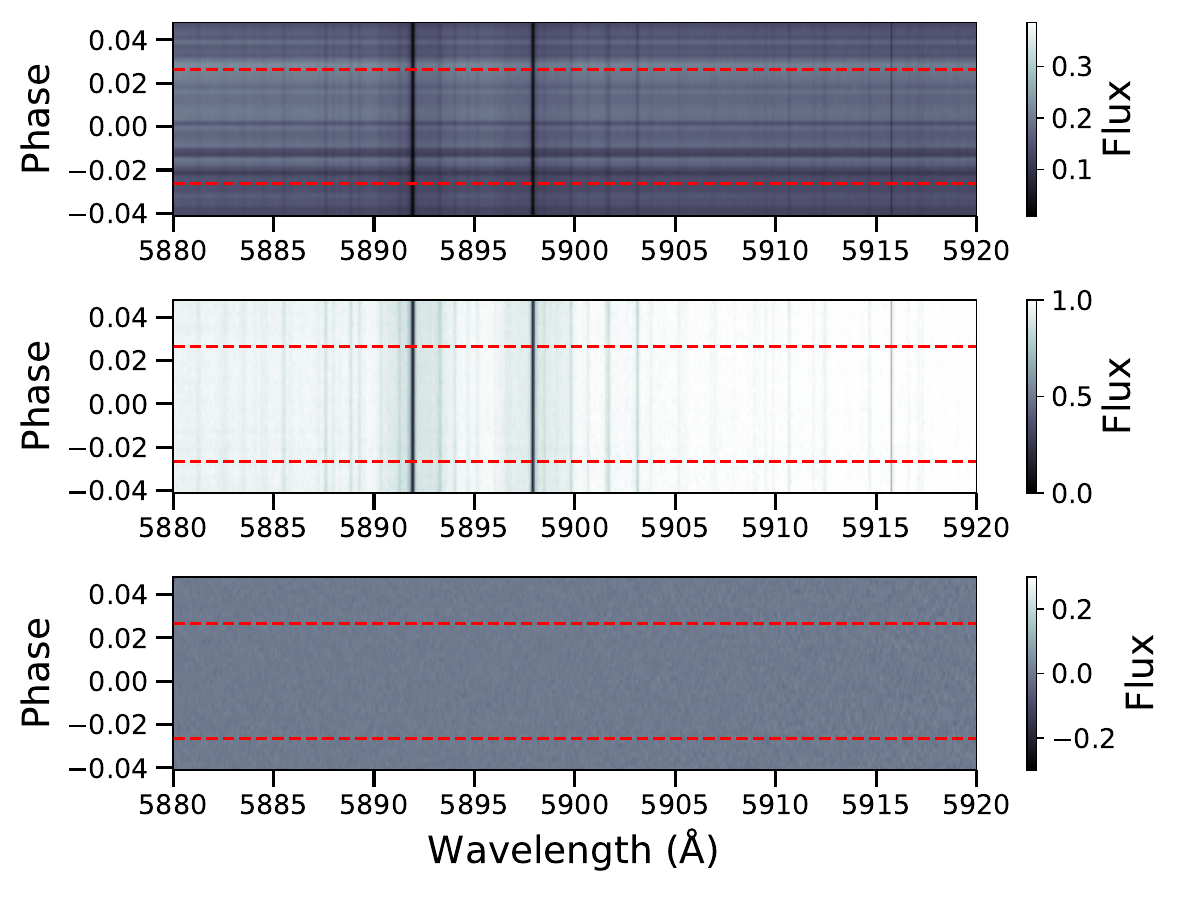}
    \caption{{\it Top panel:} The raw spectrum matrix around the Na\,{\sc i} doublet. {\it Middle panel:} The normalized spectrum matrix after pre-processing (see Section~\ref{spec_pre_processing} for details). {\it Bottom panel:} The spectrum matrix after five \code{SYSREM} iterations. In all panels, the horizontal red dashed lines represent the first and fourth contacts of the transit event.}
    \label{fig:reduction}
\end{figure}

Since the planetary atmospheric signals are typically obscured by telluric absorption and stellar contamination, we employed the \code{SYSREM} algorithm \citep{Tamuz2005,Mazeh2007} to iteratively remove these time-dependent systematic effects. We note that the Doppler-shifted planetary signals could also be erased if an excessive number of iterations is adopted, and thus it is crucial to select an optimal iteration number to maximize the planetary signal. We applied \code{SYSREM} to the spectrum matrix 8 times, and computed the difference between the residual spectrum matrix of each pair of two consecutive runs \citep{Spring2022}. We found that the standard deviation of the flux variation of all orders becomes negligible after five and four iterations for the VIS and NIR band data, respectively. Therefore, we used the residual spectra after the corresponding numbers of \code{SYSREM} runs for subsequent cross-correlation analysis. Figure~\ref{fig:reduction} illustrates the raw and normalized spectrum matrix as well as the outputs after five \code{SYSREM} iterations around the Na\,{\sc i} doublet. 

\subsubsection{Template Construction}
We constructed synthetic high-resolution transmission spectral templates by using the \code{petitRADTRANS} code \citep{Molliere2019}, which has been used in other studies \citep[e.g.,][]{Sanchez2019,Casasayas-Barris2022,Hood2024,Landman2024,Petz2024,Yang2024,Pelletier2025,Simonnin2025}. We assumed solar abundances for all species and an isothermal atmosphere with temperature around 2500 K, close to the planetary equilibrium temperature, between pressures of $10^{2}$ to $10^{-8}$ mbar. We set the continuum level to 10\,mbar since the planet atmosphere is expected to be rather opaque at a pressure of about 3.5 mbar due to continuum absorption by $\rm H^{-}$ \citep{Kitzmann2018,Hoeijmakers2019}. The planet radius, surface gravity and stellar size were taken from \cite{Zhou2019}, as listed in Table~\ref{param}. The line lists of all atomic species are from the Kurucz database \citep{Kurucz2018} while molecules were adopted from HITRAN \citep{ROTHMAN20134} except for VO and TiO, for which we used the results from B. Plez \citep{Molliere2019} and ExoMol \citep{Tennyson2016}. All templates were then convolved with a Gaussian function to match the spectral resolution of the observation, and interpolated to the CARMENES wavelength stamps. 


\subsubsection{Cross Correlations with Templates}\label{cc_results}

For each investigated species, we cross-correlated the residual spectra with the template using the \code{crosscorrRV} algorithm implemented in the \code{PyAstronomy} package \citep{Czesla2019}. In short, the template was Doppler-shifted by a list of velocities $v$ and the cross-correlation coefficients are then determined via 
\begin{equation}
    CC(v, \phi) = \sum^{N}_{i=1}f_{i}(\phi)\times T_i(v),
\end{equation}
where $f_{i}(\phi)$ is the flux of the $i_{\rm th}$ pixel of the spectrum obtained at orbital phase $\phi$, and $T_i(v)$ represents the template with wavelength shifted by a velocity of $v$. We carried out the calculation in the terrestrial rest frame and looped for all spectral orders with test radial velocities ranging from $-200$ to $200$ $\rm km\ s^{-1}$ stepped by $\rm 0.8\ km\ s^{-1}$. For each target species, we stacked the cross-correlation maps of all spectral orders, except for the bluest four orders of the VIS data, which have low S/N and the three orders (1345--1435 \AA) of the NIR data containing water lines that have strong telluric absorptions.

After obtaining the stacked cross-correlation map, we transformed the results to the planet rest frame based on the planet velocity $v_{p}$:
\begin{equation}
    v_{p}(t,K_p) = v_{\rm sys} - v_{\rm bary}(t) + K_p\sin2\pi\phi(t),
\end{equation}
where $v_{\rm sys}$ is the systemic velocity of the star, $v_{\rm bary}$ is the barycentric velocity, $K_p$ is the semi-amplitude of the planetary radial velocity and $\phi(t)$ is the orbital phase at time stamp $t$. Despite having a prior estimate ($K_p\approx{\rm 187\ km\ s^{-1}}$), we performed an agnostic search and shifted the map to a linearly-spaced grid of $K_p$ values from 0 to 500 $\rm km\ s^{-1}$ in steps of $\rm 1\ km\ s^{-1}$. We finally co-added the in-transit cross-correlation results and constructed a $K_p$-$\Delta v$ map. All cross-correlation coefficients were converted to S/N through dividing by the standard deviation outside the region of $-40 \leq \Delta v\leq 40$ ${\rm km\ s^{-1}}$ and $100 \leq K_p\leq 250$ ${\rm km\ s^{-1}}$, in which the signal is expected to be located \citep{Brogi2018}. 

Generally, we anticipate an absorption signal to appear near the predicted $K_p$, and $\Delta v$ around $\rm 0\ km\ s^{-1}$. Nevertheless, the location of the signal may have several $\rm km\ s^{-1}$ offsets due to additional motions that are not taken into account such as atmospheric winds \citep{Kempton2012}, manifesting themselves as a blue or red shift along the $\Delta v$ axis \citep{Snellen2010,Wyttenbach2015,Louden2015,Yan2022}. The planetary winds are supposed to be driven by the large temperature gradient from the dayside to the nightside, which is more significant if the planet has higher insolation as in the case of UHJs \citep{Komacek2016}. Therefore, to claim a robust detection, we required that the signal has a cross-correlation S/N over 4 and it is located close to the expected position ($K_p=187\ {\rm km\ s^{-1}}$ and $\Delta v=0\ {\rm km\ s^{-1}}$), within $\delta K_{p}=40\ {\rm km\ s^{-1}}$ as well as $\delta \Delta v=25\ {\rm km\ s^{-1}}$. Moreover, we also defined a tentative detection if the signal satisfies the same $K_p$ and $\Delta v$ thresholds but with S/N between 3 and 4. However, when there are more than 5 peaks with $\rm 3<S/N<4$ in the cross-correlation map, even if one of them is close to the expected location, we attributed the signal to random noise. We located the signal with the maximum S/N in the $K_p$-$\Delta v$ map, and fitted a one-dimensional Gaussian profile (a standard Gaussian function plus an offset) to the cross-correlation result. For each species, we generated a three-column diagnostic plot including the 2D cross-correlation residual map, the 2D $K_p$-$\Delta v$ map and the 1D cross-correlation function, as shown in Figure~\ref{fig:robust_detection}. 



\section{Results}\label{section_results}

\subsection{Results from Single-Line Analysis}

\begin{table*}
\centering
\caption{
\label{table - Results TS}
Results from the single line transmission spectroscopy analyses.
}
\centering
\begin{tabular}{lcccccc}
\hline \hline 
\noalign{\smallskip} 

Line & Wavelength [\AA] & Depth (\%) & EW [m\AA] & $\Delta v$ [km\,s$^{-1}$] & $\sigma$ [\AA] & FWHM [\AA] \vspace{0.05cm}\\
\hline
\noalign{\smallskip}

H$\alpha$ & 6564.70 $\pm$ 0.03  & 1.50 $\pm$ 0.13  &  11.8 $\pm$ 1.0  &  2.5 $\pm$ 1.5  &  0.32 $\pm$ 0.03  & 0.74 $\pm$ 0.07 \\

He\,{\sc i}\,NIR triplet & ... & $<$0.8 & $<$8.5  &   &  &   \\

Na\,{\sc i}\,D2 & 5891.582 $\pm$ 0.008  & 0.903 $\pm$ 0.025  &  2.6 $\pm$ 0.12  & -0.1 $\pm$ 0.4  & 0.116 $\pm$ 0.005 &  0.274 $\pm$ 0.012  \\
Na\,{\sc i}\,D1 & ... & $\lesssim$0.7  &   &   &  &   \\

Li\,{\sc i} & ... & $<$0.7  &   &   &  &   \\

K\,{\sc i} & 7701.28 $\pm$ 0.10 & 0.41$^{+0.06}_{-0.04}$  & 4.8$^{+1.3}_{-1.2}$  &  7 $\pm$ 4  &  0.46$^{+0.13}_{-0.12}$  &  1.10$^{+0.33}_{-0.30}$   \\

Ca\,{\sc ii}\,IRT$_1$ & 8500.37 $\pm$ 0.06  & 0.68 $\pm$ 0.17  & 3.0 $\pm$ 0.7  &  0.9 $\pm$ 2.3  &  0.17$^{+0.05}_{-0.04}$  &  0.42$^{+0.11}_{-0.10}$ \\
Ca\,{\sc ii}\,IRT$_2$ & 8544.54 $\pm$ 0.05  & 1.13$^{+0.21}_{-0.18}$  &  6.7$^{+1.2}_{-1.1}$    & 3.5$^{+1.8}_{-1.6}$  &  0.24$^{+0.06}_{-0.05}$  &  0.55$^{+0.13}_{-0.11}$ \\
Ca\,{\sc ii}\,IRT$_3$ & 8664.67$^{+0.12}_{-0.09}$   & 0.75$^{+0.18}_{-0.16}$  & 5.3$^{+1.4}_{-1.3}$  & 5.1$^{+4.0}_{-3.0}$  &  0.29$^{+0.08}_{-0.07}$   &  0.67$^{+0.20}_{-0.17}$  \\

\noalign{\smallskip}
\hline
\end{tabular}
\end{table*}

\begin{figure*}
    \centering
\includegraphics[width=\linewidth]{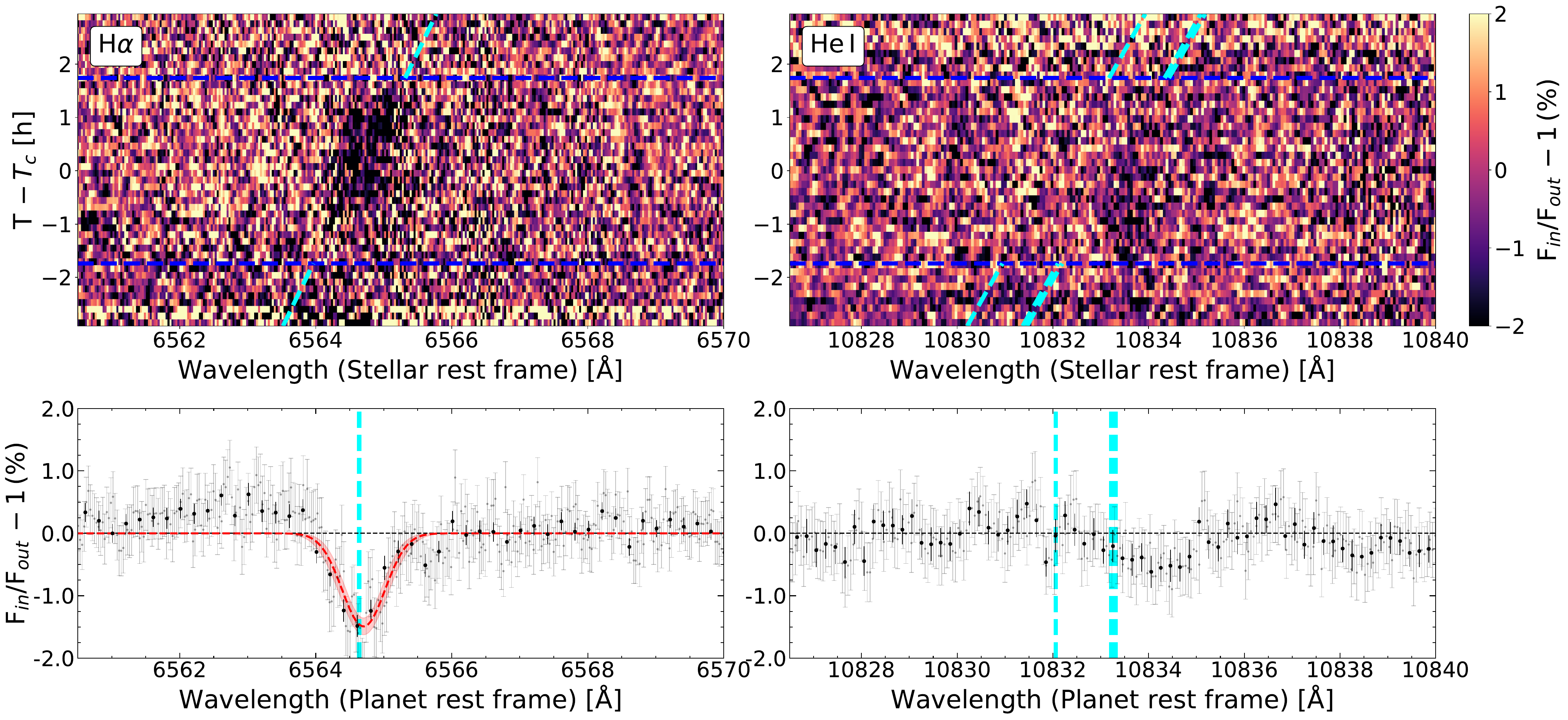}
    \caption{\label{Fig: TS Halpha Helium} Residual maps and transmission spectra around H$\alpha$ (\textit{left}) and the He\,{\sc i} NIR triplet (\textit{right}) lines. \textit{Top panels}: Residual maps in the stellar rest frame. The time since mid-transit time ($T_c$) is shown on the vertical axis, wavelength is on the horizontal axis, and relative absorption is color-coded. Dashed blue horizontal lines indicate the transit duration. Dashed cyan tilted lines show the theoretical trace of the planetary signals. \textit{Bottom panels}: Transmission spectra obtained combining all the spectra between the first and fourth contacts. We show the original data in light gray and the data binned by 0.2\,\AA\ in black. The best Gaussian fit model is shown in red along with its $1\sigma$ uncertainties (shaded red region). Dashed cyan vertical lines indicate the H$\alpha$ (\textit{left}) and the He\,{\sc i} triplet (\textit{right}) lines positions. All wavelengths in this figure are given in vacuum.
    }
\end{figure*}

\begin{figure*}
    \centering
    \includegraphics[width=\linewidth]{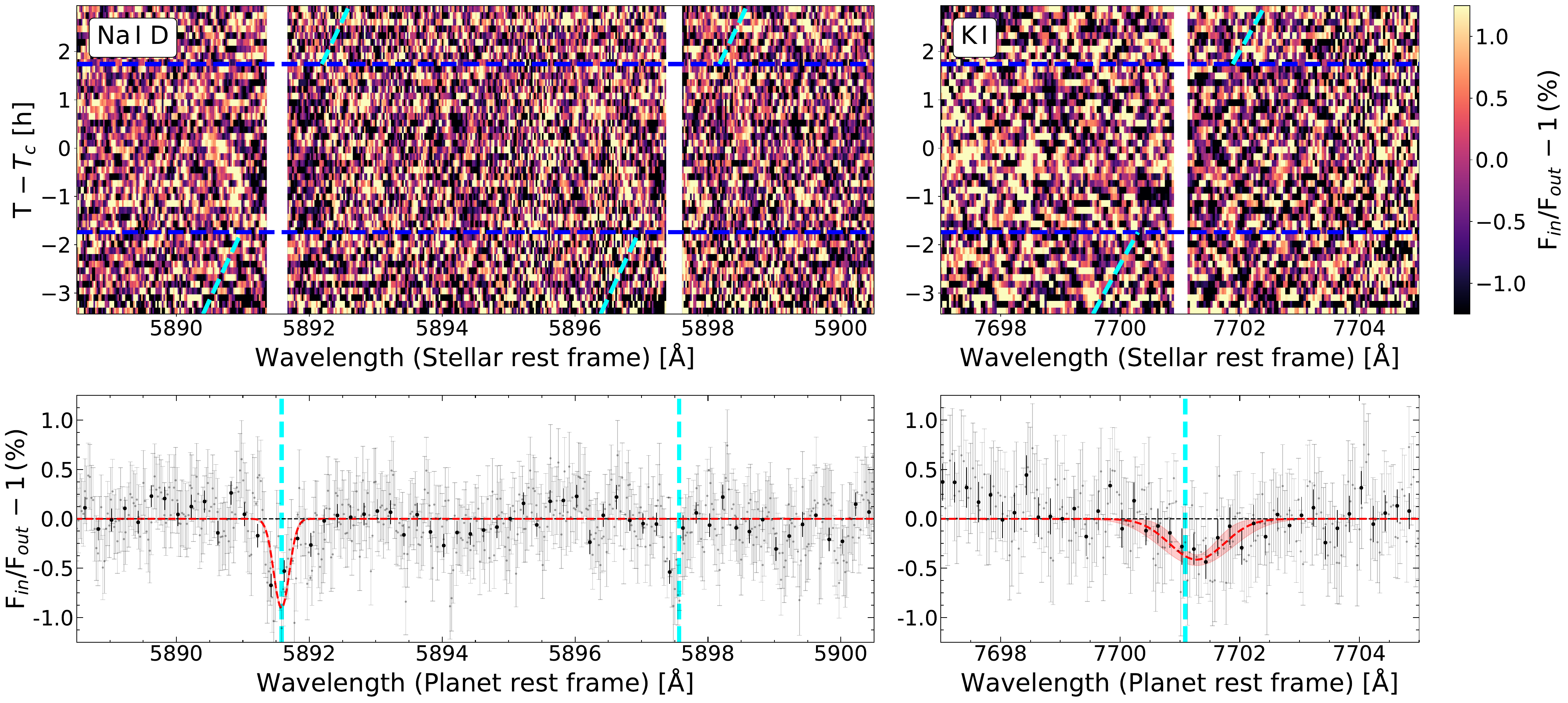}
    \caption{\label{Fig: TS Na Doublet} Same as Figure\,\ref{Fig: TS Halpha Helium} but for the Na\,{\sc i} doublet lines (left panels) and K\,{\sc i} $\lambda$7701\angstrom\ (right panels). The white vertical bands are the masked regions due to interstellar medium absorption. Only a fit for Na\,{\sc i}\,D2 is presented. A shallow dip can be seen in the result of K\,{\sc i}, which might be due to atmospheric absorption.
    }
\end{figure*}

\begin{figure*}[]
    \centering
    \includegraphics[width=\linewidth]{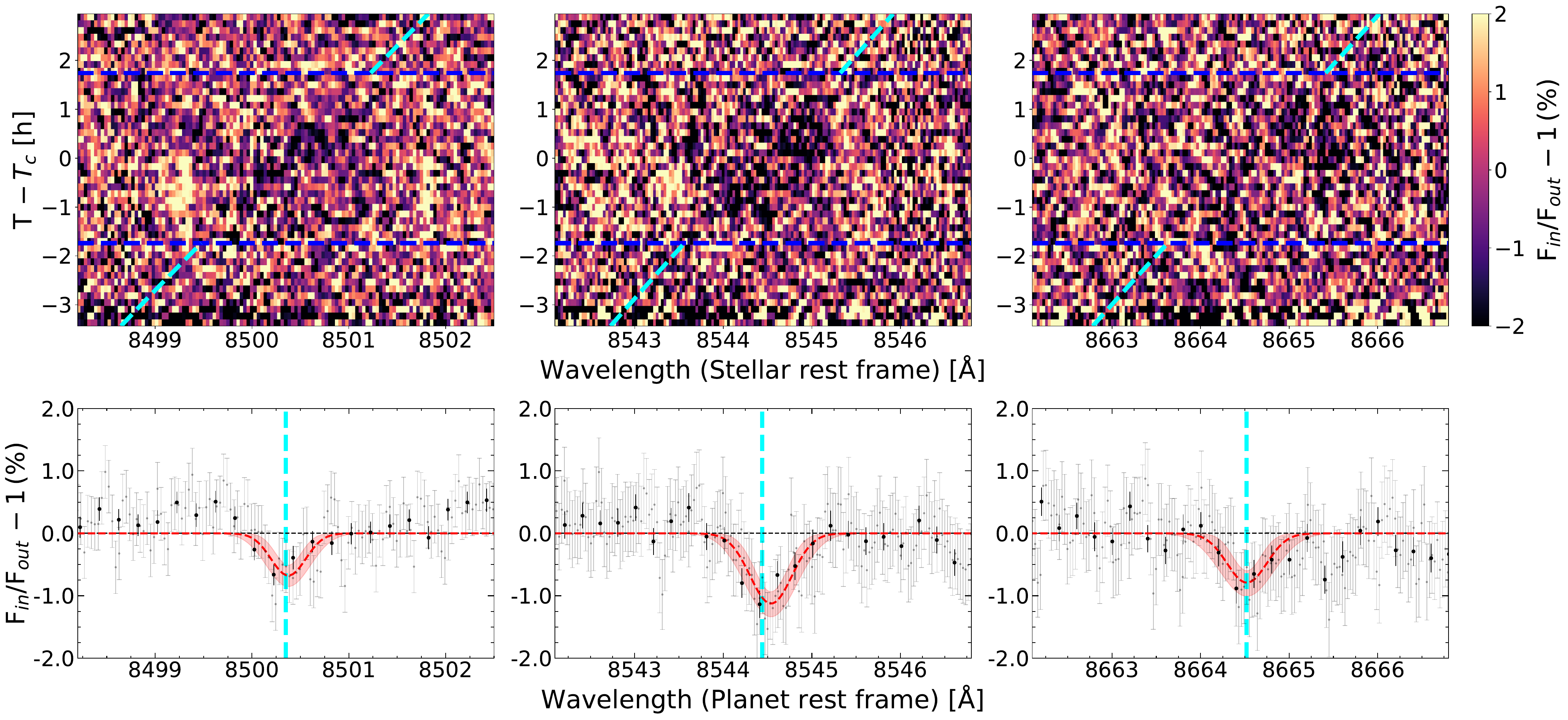}
    \caption{\label{Fig: TS Ca II} Same as Figure\,\ref{Fig: TS Halpha Helium} but for the Ca\,{\sc ii}\,IRT lines: $\lambda$8500\,\angstrom\ (left), $\lambda$8544\,\angstrom\ (middle), $\lambda$8664\,\angstrom\ (right). 
    }
\end{figure*}

Figures\,\ref{Fig: TS Halpha Helium}, \ref{Fig: TS Na Doublet}, \ref{Fig: TS Ca II}, and \ref{Fig: TS Li K} show the residual maps and TS for the single lines inspected. The extracted information from the TS is summarized in Table\,\ref{table - Results TS}. Overall, our CARMENES observations confirm previous single-line detections reported for HAT-P-70\,b by \cite{BelloArufe2022} and \cite{Langeveld2025}. For the first time, we provide constraints on the He triplet and find a tentative signal of K\,{\sc i}.

We find a significant absorption of H$\alpha$, visible in the residual map and TS (see Figure~\ref{Fig: TS Halpha Helium} left panels). Although the RM+CLV effects were corrected, the TS shows a feature to the left of the planet absorption that suggests these effects may have not been fully removed. This feature is not detected on the other lines. The 1.50$\pm$0.13\,\% signal detected with CARMENES is in excellent agreement with that reported by \citet[depth = 1.56$\pm$0.15\,\%]{BelloArufe2022} from TNG/HARPS-N.
We also explored the atmospheric evaporation of HAT-P-70\,b through the He\,{\sc i} NIR triplet accessible by CARMENES NIR. The residual map and TS shown in Figure~\ref{Fig: TS Halpha Helium} (right panels) show a broad absorption feature which is significantly red-shifted and looks vertical in the residual map. This signal comes from small OH telluric residuals and not from the atmosphere of the planet. From the scattering of the TS, we derive a 3$\sigma$ upper limit of 0.8\,\% for a He planetary absorption.

The stellar Na\,{\sc i} doublet lines (D1: 5897\,\AA, D2: 5891\,\AA) are affected by a strong interstellar medium (ISM) absorption.
We masked the central region of the lines because the ISM absorption produces artifacts when subtracting the stellar reference spectrum, which affects the TS. We find a significant absorption of Na\,{\sc i}\,D2 of $\sim$0.9\,\%. Although the residual map and TS (Figure~\ref{Fig: TS Na Doublet}) present a tentative Na\,{\sc i}\,D1 absorption signal, we are not able to fit this narrow feature with a Gaussian profile. For this line, we report the binned point value at the line position as the upper limit because it is comparable to the original TS depth.
In general terms, our D2 detection and D1 upper limit are consistent with the Na\,{\sc i} doublet detection reported by \cite{BelloArufe2022}.

Using Gemini South/GHOST, \cite{Langeveld2025} measured significant planetary absorption from the Ca\,{\sc ii}\,IRT lines at 8500\,\AA, 8544\,\AA, and 8664\,\AA. Figure\,\ref{Fig: TS Ca II} presents the residual map and TS from CARMENES. We detect significant absorption for the three lines of 0.7\,\%, 1.1\,\%, and 0.75\%, respectively. The absorption we measure is in agreement with that reported by \cite{Langeveld2025}.

We also explored the possible planetary absorption of Li\,{\sc i} $\lambda$6710\,\AA\ (Figure\,\ref{Fig: TS Li K}) and K\,{\sc i} $\lambda$7701\,\AA\ (Figure\,\ref{Fig: TS Na Doublet} right panels). The TS of the Li line is mainly flat and we can only place a 0.7\,\% upper limit from the scattering of the TS. We masked a narrow deep absorption line likely from K ISM absorption close to the K\,{\sc i} line \citep{K_ISM}. The TS of the K line shows a broad $\sim$0.4\% absorption feature, but the planetary trace is not clear in the residual map. The signal is fitted with significance, and it could match with the tentative detection of K\,{\sc i} that we find in cross-correlation (Figure\,\ref{fig:tentative_detection}).

Previous results from \cite{BelloArufe2022} and \cite{Langeveld2025} show that both CC and single-line results are slightly blue-shifted ($\Delta v$ $<$ 0\,${\rm km\ s^{-1}}$). However, here we obtained null to slightly red-shifted absorptions. Although the RM+CLV effects are corrected, there is some noise in the residual maps that could be related to these effects (see Figure~\ref{Fig: TS Ca II}). These residuals might affect the TS, deforming the planetary signal and resulting in red-shifted absorptions.

\subsection{Results from Cross-Correlation Analysis}

\begin{figure*}
    \centering
    \includegraphics[width=\textwidth]{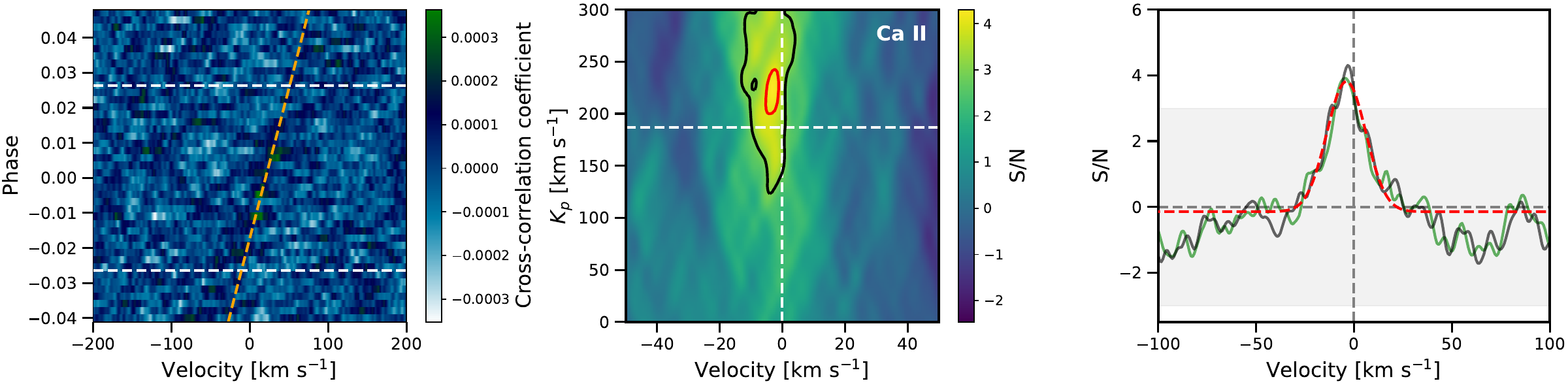}
    \includegraphics[width=\textwidth]{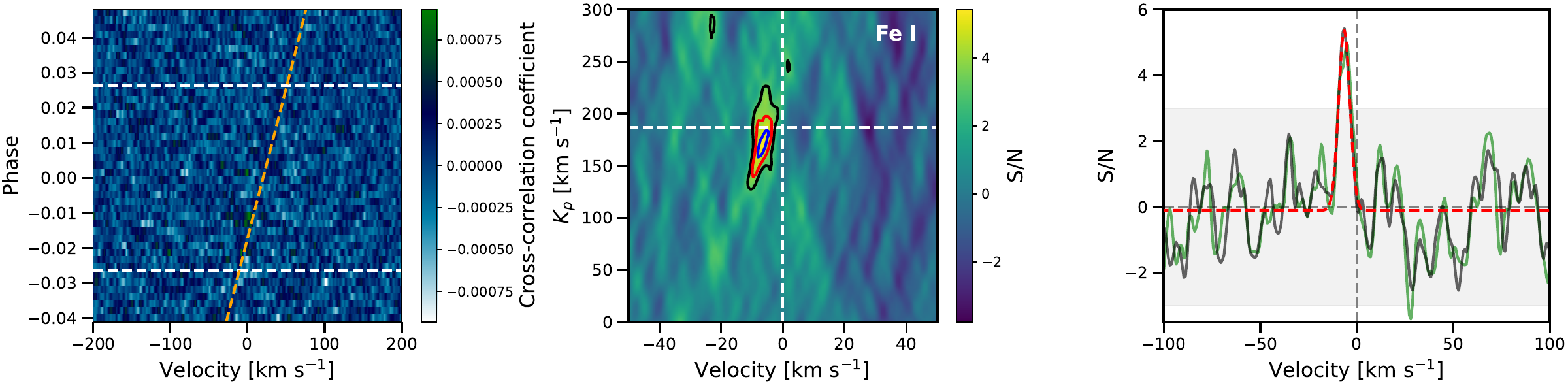}
    \caption{Cross-correlation results of \tar b with robust detections that have $\rm S/N\geq 4$ (top: Ca\,{\sc ii}, bottom: Fe\,{\sc i}). {\it Left panel:} The cross-correlation residual map. The horizontal white dashed lines mark the first and fourth contacts of the transit event. The orange tilted dashed line represents the planet trajectory. {\it Middle panel:} The cross-correlation S/N as a function of velocity shift ($\Delta v$) and planet Keplerian velocity ($K_p$). The vertical and horizontal white dashed lines show the expected location of the planet signal ($\rm 0\,km\ s^{-1}$ and $\rm 187\,km\ s^{-1}$). The black, red and blue contour lines mark the region with S/N equal to 3, 4 and 5. {\it Right panel:} The 1D cross-correlation function at the best $K_p$ (black) along with the best-fit Gaussian profile (red). The green line is the same result but based on the literature $K_p$ (see Section~\ref{cc_results} for details). The gray shaded region highlights the cross-correlation S/N between -3 and 3.}
    \label{fig:robust_detection}
\end{figure*}

\begin{figure*}
    \centering
    \includegraphics[width=\textwidth]{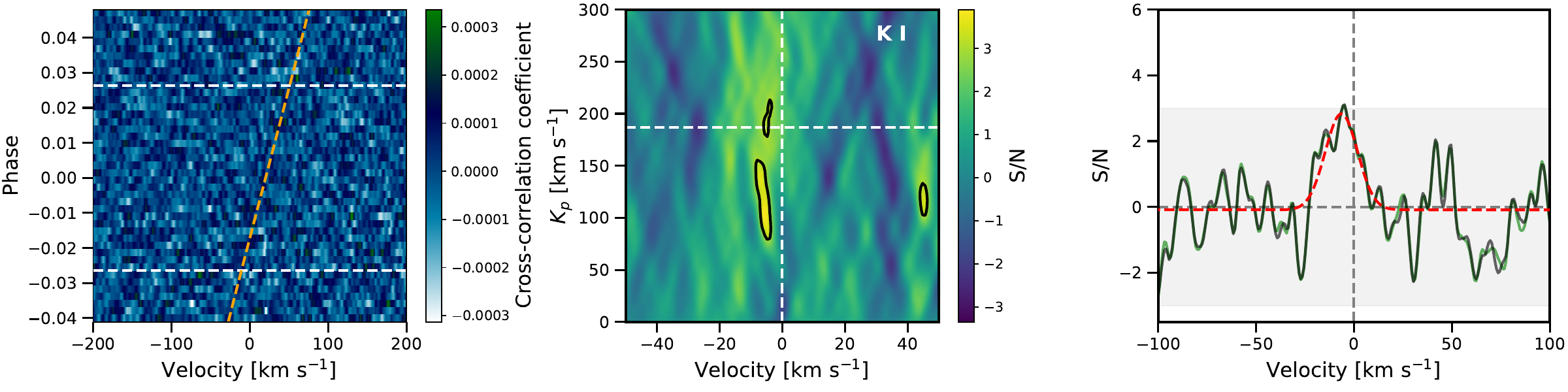}
    \caption{Same as Figure~\ref{fig:robust_detection} but for the tentative detection of K\,{\sc i} with $\rm 3\leq S/N<4$.}
    \label{fig:tentative_detection}
\end{figure*}

Figures~\ref{fig:robust_detection} and \ref{fig:tentative_detection} present the robust and the tentative detections from the cross-correlation analysis. We summarize the details in Table~\ref{ccresults}. Overall, we verify previous detections of Ca\,{\sc ii} and Fe\,{\sc i} by \cite{BelloArufe2022} and report the new tentative detection of K\,{\sc i}. However, we note that \cite{BelloArufe2022} identified significant absorption of several other atomic species such as Cr\,{\sc i}, Fe\,{\sc ii} and V\,{\sc i} using HARPS-N whereas they are not detected in the CARMENES data. The discrepancy is mainly due to the following two reasons: the lower spectral S/N of the CARMENES data limit the detection sensitivity; the wavelength coverage of CARMENES does not extend below 5000\,\AA, where most metal absorption lines are located.

\subsubsection{Detections of Ca\,{\sc ii} and Fe\,{\sc i}}

We identified a strong and extended Ca\,{\sc ii} absorption signature with a maximum S/N of 4.3 in the $K_p$-$\Delta v$ map, peaking at $K_p=222\pm24\ {\rm km\ s^{-1}}$ and blue-shifted by about $-3.2\,\pm\,2.5\,{\rm km\ s^{-1}}$, in agreement with the previous results of $K_p=236^{+74}_{-163}\ {\rm km\ s^{-1}}$ and $\Delta v=-7^{+15}_{-5}\ {\rm km\ s^{-1}}$ within $1\sigma$ \citep{BelloArufe2022}. Unlike HARPS-N that covers the Ca\,{\sc ii} H\&K lines at 3935 and 3970 $\angstrom$, we confirmed the signal independently based on its near-infrared triplet at 8500, 8544, and 8664 $\angstrom$, similar to \cite{Langeveld2025}. Through fitting a Gaussian profile at the best $K_p$, we obtained a FWHM of $24.3\pm1.5\ {\rm km\ s^{-1}}$ by transforming the standard deviation $\sigma$ of the Gaussian function using FWHM=2.355$\times \sigma$, much narrower than the $31.5\pm1.9\ {\rm km\ s^{-1}}$ found by \cite{BelloArufe2022}. The 1D cross-correlation function peaks at $\Delta v=-3.4\pm0.6\ {\rm km\ s^{-1}}$, suggesting the presence of winds in the atmosphere. 

The Fe\,{\sc i} absorption exhibits a similar blueshift ($\Delta v=-6.4\ {\rm km\ s^{-1}}$), appearing at $K_p=169\pm15\ {\rm km\ s^{-1}}$ with a maximum S/N of 5.4. The measurements agree well with the findings of $K_p=169^{+21}_{-18}\ {\rm km\ s^{-1}}$ and $\Delta v=-6^{+1}_{-2}\ {\rm km\ s^{-1}}$ from \cite{BelloArufe2022}. A Gaussian profile fitting at the optimal $K_p$ yields a S/N of $5.48\pm0.46$, a $\Delta v$ of $-6.51\pm0.28\ {\rm km\ s^{-1}}$ and a FWHM of $6.92\pm0.68\ {\rm km\ s^{-1}}$.


\subsubsection{Tentative and Non-Detections}

Apart from the robust detections of Ca\,{\sc ii} and Fe\,{\sc i}, we also tentatively found the presence of K\,{\sc i}, as in the single line analysis. The neutral potassium K\,{\sc i} shows a broad signal (FWHM=$19.4\pm2.2\ {\rm km\ s^{-1}}$) analogous to Ca\,{\sc ii} with a maximum S/N of about 3.1. Similarly, it is blue-shifted by a few ${\rm km\ s^{-1}}$ and it shows up at $K_p=190\ {\rm km\ s^{-1}}$, in accordance with the expected $K_p$. However, the large scatter in the 1D cross-correlation function limits the peak S/N of the best-fit Gaussian profile to a value of $2.92\pm0.28$, slightly below the threshold of S/N=3. We emphasize that this marginal signal barely meet our definition of tentative detection ($\rm 3\leq S/N<4$) and there are two other peaks in the cross-correlation map, though the total number of these signals are smaller than 5. Future observations are required to validate or rule out its existence. 



Moreover, we identified a significant absorption of vanadium oxide (VO), an indicator of thermal inversion \citep[e.g.,][]{Hubeny2003,Fortney2008,Knutson2008,Evans2016}, with a maximum S/N of 4.3 using the template from B. Plez \citep{Molliere2019}. The signal is located at $K_p = 193\,\pm\,8\ {\rm km\ s^{-1}}$ but redshifted by about $20.0\pm1.2\ {\rm km\ s^{-1}}$. We examined the outputs of the residual spectrum matrix after 3-8 \code{SYSREM} iterations where we find that the signal persists with a similar S/N. However, this signal disappears when changing to the $\rm VO$ template from ExoMol \citep{Tennyson2016}. Instead, another signal that has a S/N of about 3.4 arises at similar $K_p$ but with a much lower redshifted velocity about $4.8\pm1.0\ {\rm km\ s^{-1}}$ (Figure~\ref{fig:VO}). We note that there are multiple peaks in the $K_p$-$\Delta v$ map using the ExoMol template with a similar S/N. We refer the reader to \cite{Regt2022} for more details of the different $\rm VO$ templates. Although the cross-correlations with two different templates both result in a signal with S/N above 3, it is likely due to random noise. Therefore, we do not claim it as a tentative detection but report it as an non-conclusive result since the two signals are not consistent with each other. 


Results of other species with non-detections including the $K_p$-$\Delta v$ maps and the 1D cross-correlation functions at the literature $K_p$ are presented in Figures~\ref{fig:non_detection_ato_ion} and \ref{fig:non_detection_molecular}.


\begin{table*}
    \centering
    \caption{Results from the cross-correlation analyses. }
    \begin{tabular}{lccc|cccc}
        \hline\hline
               &\multicolumn{3}{c}{Highest S/N signal around literature $K_p$}      &\multicolumn{4}{c}{Gaussian profile at $K_p$ with highest S/N signal} \\\hline
        Species &$\rm S/N_{max}$ &$\Delta v\ [\rm km\ s^{-1}]$ &$K_p\ [\rm km\ s^{-1}]$ &S/N &$\Delta v\ [\rm km\ s^{-1}]$ &$\sigma$ $[\rm km\ s^{-1}]$ &Offset\\\hline
        \textit{Detections$^{[1]}$}: & & & & & &\\
        Ca\,{\sc ii} &$4.3$ &$-3.2\pm2.5$ &$222\pm24$ &$3.96\pm0.19$ &$-3.39\pm0.57$ &$10.35\pm0.69$ &$-0.15\pm0.03$\\
        Fe\,{\sc i} &$5.4$ &$-6.4\pm2.3$ &$169\pm15$ &$5.48\pm0.46$ &$-6.51\pm0.28$ &$2.94\pm0.29$ &$-0.10\pm0.04$\\
         \textit{Tentative Detections$^{[2]}$}: & & & & & &\\
        K\,{\sc i} &$3.1$ &$-4.8\pm1.5$ &$190\pm17$ &$2.92\pm0.28$ &$-6.41\pm0.93$ &$8.23\pm0.95$ &$-0.09\pm0.05$\\
       \textit{Non-conclusive results$^{[3]}$}: & & & & & &\\
       $\rm VO$ (B. Plez template) &$4.3$ &$20.0\pm1.2$ &$193\pm8$ &$4.08\pm0.61$ &$20.32\pm0.26$ &$1.60\pm0.26$ &$-0.04\pm0.04$\\
       $\rm VO$ (ExoMol template) &$3.4$ &$4.8\pm1.0$ &$193\pm10$ &$3.61\pm0.86$ &$5.10\pm0.26$ &$0.96\pm0.26$ &$-0.04\pm0.05$\\
         \hline
    \end{tabular}
    \begin{tablenotes}
    \item[1][1]~A detection is defined as a signal with $\rm S/N\geq 4$; [2]~A tentative detection is defined as a signal with $\rm 3\leq S/N<4$; [3]~Although the cross-correlations of $\rm VO$ using templates from B. Plez and ExoMol both lead to a signal that meets the S/N=3 boundary of a tentative detection, the two signals are not consistent with each other thus we report as a non-conclusive result. 
    \end{tablenotes}
    \label{ccresults}
\end{table*}

\section{Discussion}\label{section_discussions}

\subsection{H$\alpha$ detection is more common in young Jupiter-sized planets}

\begin{figure*}[ht!]
    \centering
    \includegraphics[width=0.9\linewidth]{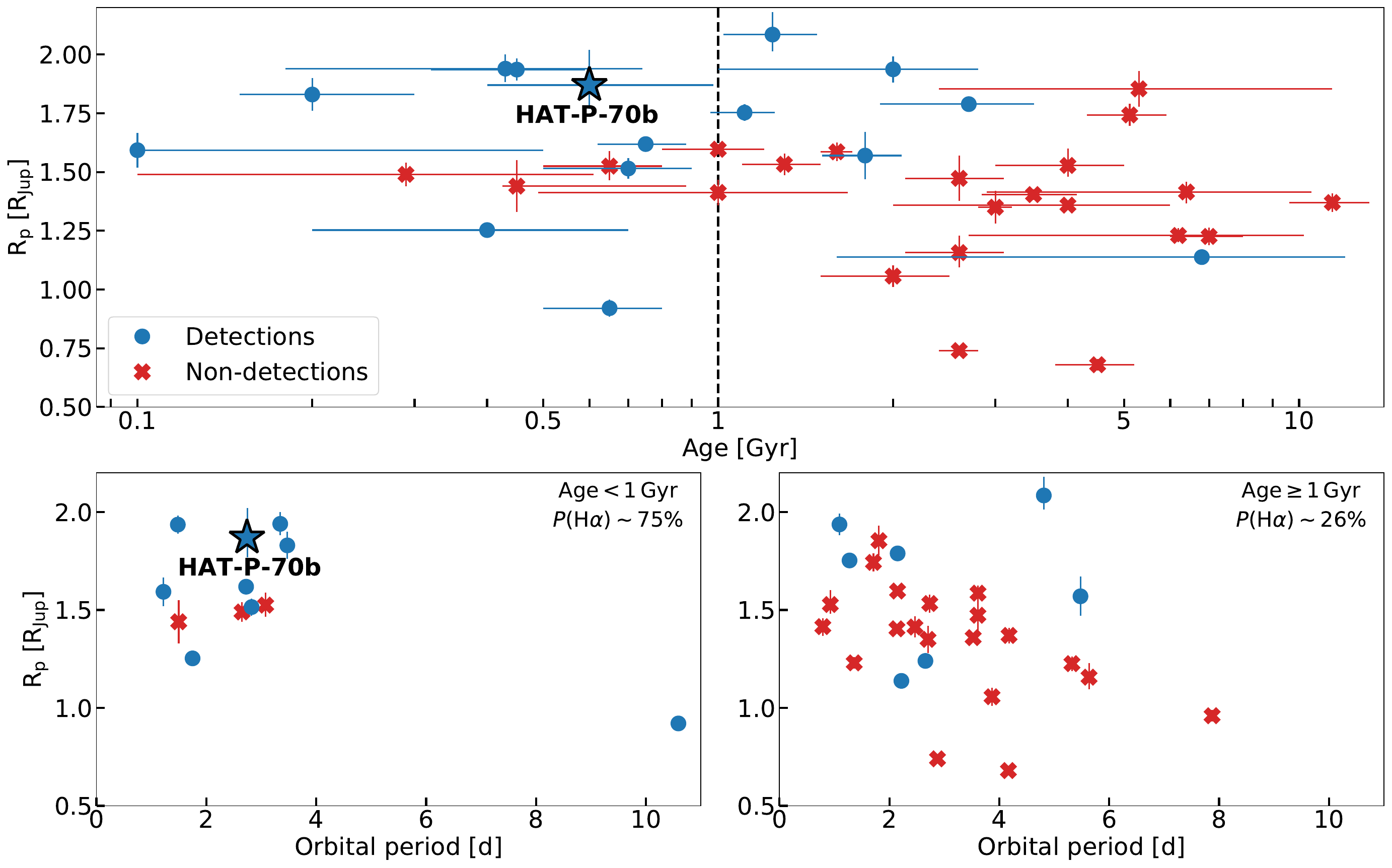}
    \caption{
    Distribution of H$\alpha$ observations for gas giants as function of stellar age vs. planetary radius (top panel), and period vs. radius diagram for $<$1\,Gyr-old (bottom left panel) and $\ge$1\,Gyr-old (bottom right panel) planets. \tar b is labeled and marked with a blue star in the panels. Planets with detection of H$\alpha$ are marked with blue dots, while those with H$\alpha$ non-detections/upper limits are marked with red crosses. We indicate the 1\,Gyr value with a dashed black horizontal line in the top panel. Results are retrieved from the ExoAtmospheres website.
    }
    \label{Fig: Age - Rp Halpha}
\end{figure*}

We explored if H$\alpha$ detections on gas giants have any correlation with planet or host star properties. We combined the result of HAT-P-70\,b with the compilation of H$\alpha$ studies available at the ExoAtmospheres\footnote{\url{https://research.iac.es/proyecto/exoatmospheres/index.php} (24 June 2025).} website, from where we adopted the classification of detections and upper limits. A detection means that the TS shows an absorption that is at least 3$\sigma$ over the noise level. The results are considered as upper limits when there is no evidence of planetary absorption. We note that the median S/N of the transmission spectroscopic observations with and without H$\alpha$ detections in the literature are similar, on average above 35. Therefore, we consider that the result is not biased to the S/N of the observation.

We found a tentative relation between H$\alpha$ detections and the age of the host stars.
Figure \ref{Fig: Age - Rp Halpha} top panel shows the stellar age vs. planetary radius for planets with radii between 0.5--2.2\,$R_{\rm Jup}$.
Detections come mainly from the planets with well defined ages below 1\,Gyr, where all except three (TOI-1431\,b, TOI-2046\,b, \citealp{Orell-Miquel2024}; and KELT-17\,b, \citealp{Stangret2022}) have positive results (from left to right in Figure\,\ref{Fig: Age - Rp Halpha} top panel: WASP-33\,b, \citealp{Yan2021_wasp33}; MASCARA-2\,b, \citealp{Mascara2_2023}; WASP-52\,b, \citealp{Chen2020}; WASP-178\,b, \citealp{Damasceno2024}; KELT-9\,b, \citealp{Yan2018_KELT-9b}; TOI-5398\,b, \citealp{DArpa2024}; MASCARA-4\,b, \citealp{Zhang2022}; and WASP-189\,b, \citealp{Prinoth2024}).

The bottom panel of Figure~\ref{Fig: Age - Rp Halpha} shows the period-radius diagrams for young ($<$1\,Gyr-old, left) and old ($\ge$1\,Gyr-old, right) planets. The comparison between both groups helps to visualize the age trend discernible in the age--radius diagram. Gas giants orbiting $<$1\,Gyr-old stars seem more likely to show absorption from H$\alpha$ than those orbiting older stars, according to Figure\,\ref{Fig: Age - Rp Halpha}. To be specific, the H$\alpha$ detection rate is about 75\% for $<$1\,Gyr-old planets while it is only around 26\% for $\ge$1\,Gyr-old planets, resulting in a relative detection probability of $P_{\rm Age<1\,Gyr}({\rm H}\alpha)/P_{\rm Age\geq1\,Gyr}({\rm H}\alpha)\sim 3$.

\begin{figure}
    \centering
    \includegraphics[width=0.99\linewidth]{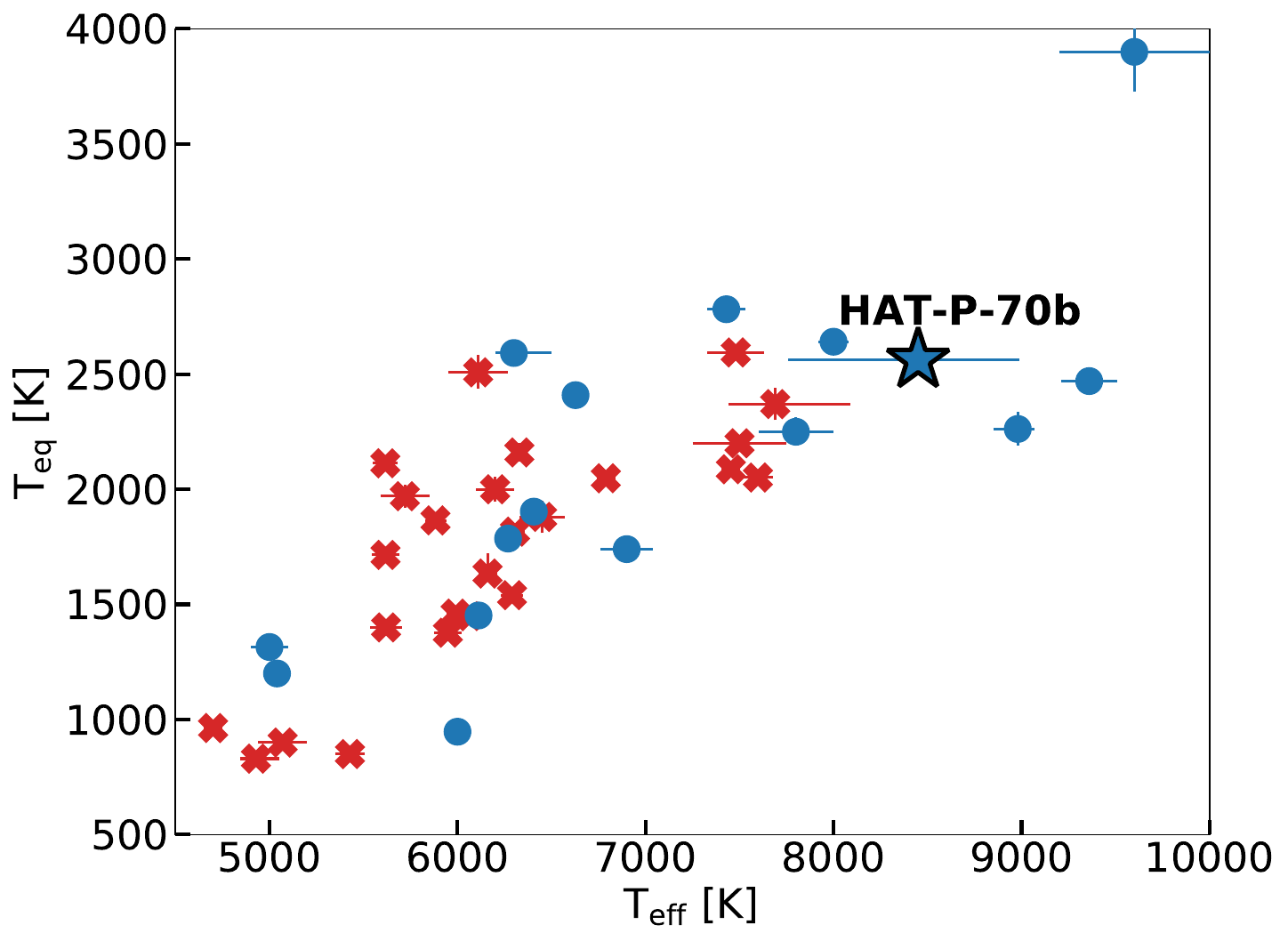}
    \caption{Effective temperature vs. equilibrium temperature for the gas giant population ($R_p$ $\gtrsim$ 0.5\,R$_{\rm Jup}$) with H$\alpha$ observations. Planets with detection of H$\alpha$ are marked with blue dots, while those with H$\alpha$ non-detections/upper limits are marked with red crosses. Results are retrieved from the ExoAtmospheres website.} \label{Fig: Teq-Teff}
\end{figure}

Theories about giant planet formation predict intense atmospheric evaporation, mainly within the first 100\,Myr (e.g., \citealp{JorgeSanz2011, Owen2013, Dawson_Johnson_2018, Gupta2020_corepower}. Thus, atmospheric absorption by H$\alpha$ is expected to be easier to detect in young planets. However, the compiled sample is consistently older than 100\,Myr, when atmospheric evaporation is predicted to be less intense \citep{Lopez2012, Owen_Jackson_2012}. Gas giants can experience mass loss without any significant consequences for their atmospheres \citep{Vissapragrada2022AJ}. Therefore, the distribution of H$\alpha$ detections for young and old gas giants from observations agrees with the theoretical predictions but it does not have evolutionary implications. However, we caution the reader that this age trend may be due to the stellar mass because most gas giants with detailed atmospheric characterizations are orbiting A and F stars, which are mostly younger than G and K stars. As shown in Figure~\ref{Fig: Teq-Teff}, H$\alpha$ detections are more common on planets around stars with higher effective temperatures. As a large gas giant orbiting a young host star, \tar b follows the general trends depicted by Figure \ref{Fig: Age - Rp Halpha} and predicted by atmospheric evaporation models.


\subsection{Constraints on mass loss rate and planetary He signal from XUV flux} 

To test the expected He\,{\sc i}~$\lambda$10833\,{\AA} triplet absorption signal in HAT-P-70b, we measured the X-ray emission of its host star. The star was observed on UT 2024 February 14 with \textit{XMM-Newton} (P.I. Corrales) for an average European Photon Imaging Camera (EPIC) exposure of 38.6\,ks \citep[sensitivity range 0.1--15 keV and 0.2--10~keV, respectively, $E/\Delta E\sim$20--50,][]{tur01,str01}. The data were reduced using the standard tools in the Science Analysis Software (\code{SAS}) v21.0, and were simultaneously fitted using the Interactive Spectral Interpretation System \citep[\code{ISIS},][]{isis} package, with a single-temperature coronal component of $\log T $(K)$ =6.96^{+0.07}_{-0.06}$, $\log EM $(cm$^{-3}$)$ =51.39^{+0.07}_{-0.08}$). An ISM absorption of $N_{\rm H} = 1.2 \times 10^{21}$\,cm$^{-3}$ was adopted according to the ISM HEASARC tool and the map from \cite{HI4PI}. The coronal abundance was fixed to the photospheric value [Fe/H]=$-$0.06\,dex. The spectral fit gives an unabsorbed X-ray luminosity $\log L_{\rm X} ({\rm erg\,s^{-1}})=28.90$, which implies a value of $\log L_{\rm X}/L_{\rm bol}=-5.78$. Since A dwarfs do not usually display X-ray emission, this low value is not unexpected for a young A3\,V star. Assuming that the X-ray emission is not contaminated by any other source, we can set a value of the expected He\,{\sc i} triplet emission based on the stellar XUV  (5--504~\AA) irradiation of the planet. Using the relation between X-rays (5--100\,\AA) and EUV (100--504\,\AA) emission, and between the XUV and He\,{\sc i}~$\lambda$10833\,{\AA} triplet equivalent width as reported in \citet{san25}, we expect a minimum value of $EW$=$2.0^{+0.5}_{-0.4}$\,m\AA, adopting the mass upper limit of the planet \citep[$M_{\rm p} <6.78$\,M$_{\rm Jup}$]{Zhou2019}. Following \cite{JorgeSanz2011}, by fitting a piecewise power-law mass-radius relation to confirmed giant planets ($R_{\rm p}>0.7$\,R$_{\rm Jup}$) with known mass and
radius\footnote{$M_{\rm p}=1.67 \cdot \log (R_{\rm
p})-0.31$ for $R_{\rm p}>0.7$\,R$_{\rm Jup}$, based on \url{exoplanet.eu} data (21 October 2025).}, we estimated that a mass of $\sim1.3$~M$_{\rm Jup}$ is more likely for \tar b. We thus obtained $EW$=$4.7^{+0.55}_{-0.49}$\,m\AA, agreeing with the upper limit set by our observations ($<8.5$\,m\AA).


It is possible to obtain an estimation of the energy-limited mass loss rate of the planet \tar b based on the XUV (5--920~\AA) irradiation of the planet \citep[][and references therein]{JorgeSanz2011}. We used the relations in \citet{san25} to get the EUV flux in the range 100--920~\AA, $\log L_{\rm EUV}$(erg\,s$^{-1}$)=29.75. This results in an expected mass loss greater than $4.1\times 10^{11}$\,g\,s$^{-1}$ (or 2.2\,M$_\oplus$\,Gyr$^{-1}$), depending on the actual planet mass.


\subsection{Ultra-Hot Jupiters with Ca\,{\sc ii} Are Also Likely to Have Fe\,{\sc i}}

At present, Fe\,{\sc i} and Ca\,{\sc ii} have been detected in the atmospheres of about a dozen UHJ planets \citep[e.g.,][]{Hoeijmakers2018,Yan2019,Stangret2020,Casasayas-Barris2022,Seidel2023}. In these studies, a detection is mostly defined as either $\rm S/N>4$ if using the cross-correlation method or $>3\sigma$ significance if using the single line analysis. We emphasize that these results were based on various instruments, hence they have different detection sensitivities given their different wavelength coverages and spectral resolutions.

\begin{figure}
    \centering
    \includegraphics[width=0.9\linewidth]{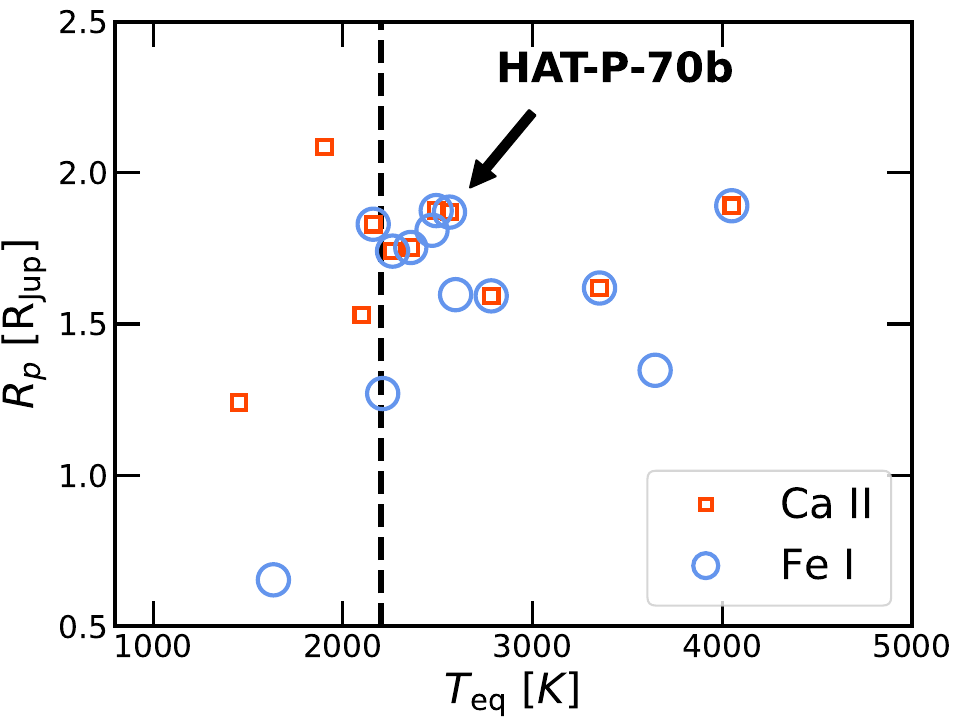}
    \caption{The equilibrium temperature and planet radius distribution of planets with either Fe\,{\sc i} (blue circle) or Ca\,{\sc ii} (red square) detections in their atmospheres. The vertical black dashed line represents the lower temperature threshold (2200K) of UHJs. \tar b is marked by a black arrow.}
    \label{fig:FeI_CaII}
\end{figure}

Figure~\ref{fig:FeI_CaII} illustrates the planet radius and equilibrium temperature distribution of giant planets that have atmospheric either Ca\,{\sc ii} or Fe\,{\sc i}, retrieved from the ExoAtmospheres data archive. Both species usually appear in hot inflated Jupiters with $T_{\rm eq}>1500$~K, as metals are likely to be condensed when the equilibrium temperatures are lower. Regarding UHJs with $T_{\rm eq}\geq 2200$~K \citep{Stangret2022,Maiz2024}, there is a tendency that UHJs have Ca\,{\sc ii} accompanied by Fe\,{\sc i} in their atmospheres. If the atmosphere of an UHJ has Ca\,{\sc ii} already, we find that the conditional probability that it also contains Fe\,{\sc i} (i.e., $P_{T_{\rm eq}\geq2200~{\rm K}}{\rm (Fe~I|Ca~II)}$) is 100\%, according to all detections known so far. In contrast, for cooler giant planets with $T_{\rm eq}\leq 2200$\,K, the presences of Ca\,{\sc ii} and Fe\,{\sc i} seem to have no correlation, though the sample size is small which impedes firm conclusions. 

The correlation between the presence of Ca\,{\sc ii} and Fe\,{\sc i} is perhaps simply due to the fact that both species are easy to detect. The Ca\,{\sc ii} has strong absorption lines while the Fe\,{\sc i} has numerous lines, both resulting in a high detection sensitivity. Nevertheless, it is possible that there are some intrinsic underlying physics between the appearances of different species. Given the high-temperature environment of UHJs, the atomic species, in either neutral or ionized states, are expected to be abundant \citep{Parmentier2018,Helling2019,Stangret2022}. However, a single transmission spectroscopic observation is generally not sufficient for a full characterization of the planetary atmosphere. If these kind of correlations are physical, one might be able to predict the existence of some elements based on the available known detections. Such priors and knowledge to the temperature-pressure profile of the planet will motivate additional spectroscopic observations using instruments covering different wavelength ranges or the same facility but aiming for higher S/N data.


\section{Conclusions}\label{section_conclusions}

In this work, we investigate the atmosphere of \tar b, an UHJ orbiting an A star, with CARMENES high-resolution transmission spectroscopy. We revisit several neutral and ionized species examined by \cite{BelloArufe2022}, and extend the study to elemental absorptions in the red and near-infrared bands. 

Using single-line analysis, we confirm the detections of H$\alpha$, Na\,{\sc i}, and Ca\,{\sc ii}\,IRT. Moreover, for the first time, we report a tentative absorption of K\,{\sc i} and constrain the atmospheric evaporation through the He triplet on this planet. Through cross-correlations, we detect the absorption of Fe\,{\sc i} and Ca\,{\sc ii} at the S/N level of about 5.4 and 4.3. Both signals are slightly blue shifted with velocities comparable to the values in the previous work \citep{BelloArufe2022,Langeveld2025}. In addition, we find the tentative detections of K\,{\sc i} with S/N barely higher than 3, requiring future observations to validate or rule out the signal. 

Putting \tar b in broader context, we find that H$\alpha$ absorption is more likely to be found in the atmospheres of inflated giant planets around stars younger than 1~Gyr. UHJs showing Ca\,{\sc ii} absorption are likely also show detectable Fe\,{\sc i} with a conditional probability $P_{T_{\rm eq}\geq2200{\rm K}}{\rm (Fe~I|Ca~II)}$ close to 100\%, according to the current observations. 

\section{Acknowledgments}

   We thank the anonymous referee for the constructive
comments that improve the quality of this work. We also thank Guo Chen for useful discussions.

   We acknowledge financial support by the National Natural Science
Foundation of China through grants
No.~12133005 (T.G., S.M.) and No.~42375118 (F.Y). 

   We also acknowledge the Agencia Estatal de Investigaci\'on
(AEI/10.13039/501100011033, PID2022-137241NB-C42) of the Ministerio de Ciencia e Innovaci\'on
and the ERDF ``A way of making Europe'' through projects
   PID2022-137241NB-C4[1:4],    
   PID2021-125627OB-C3[1:2],    
   PID2022-141216NB-I00
and the Centre of Excellence ``Severo Ochoa'' and ``Mar\'ia de Maeztu''
awards to the Instituto de Astrof\'isica de Canarias (CEX2019-000920-S),
Instituto de Astrof\'isica de Andaluc\'ia (CEX2021-001131-S) and
Institut de Ci\`encies de l'Espai (CEX2020-001058-M).
This work was also co-funded by the European Union (ERC-CoG, EVAPORATOR, Grant agreement No. 101170037).


CARMENES is an instrument at the Centro Astron\'omico Hispano en
Andaluc\'ia (CAHA) at Calar Alto (Almer\'{\i}a, Spain), operated jointly by the Junta de Andaluc\'ia and the Instituto de Astrof\'isica de Andaluc\'ia (CSIC). CARMENES was funded by the Max-Planck-Gesellschaft (MPG), the Consejo Superior de Investigaciones Cient\'{\i}ficas (CSIC), the Ministerio de Econom\'ia y Competitividad (MINECO) and the European Regional Development Fund (ERDF) through projects FICTS-2011-02, ICTS-2017-07-CAHA-4, and CAHA16-CE-3978, and the members of the CARMENES Consortium (Max-Planck-Institut f\"ur Astronomie, Instituto de Astrof\'{\i}sica de Andaluc\'{\i}a, Landessternwarte K\"onigstuhl, Institut de Ci\`encies de l'Espai,
   Institut f\"ur Astrophysik G\"ottingen,
   Universidad Complutense de Madrid,
   Th\"uringer Landessternwarte Tautenburg,
   Instituto de Astrof\'{\i}sica de Canarias,
   Hamburger Sternwarte,
   Centro de Astrobiolog\'{\i}a and
   Centro Astron\'omico Hispano-Alem\'an),
   with additional contributions by the MINECO,
   the Deutsche Forschungsgemeinschaft (DFG) through the Major Research
Instrumentation Programme and Research Unit FOR2544 ``Blue Planets
around Red Stars'',
   the Klaus Tschira Stiftung,
   the states of Baden-W\"urttemberg and Niedersachsen,
   and by the Junta de Andaluc\'{\i}a.

\appendix

\section{Non-conclusive Results and Non-Detection Results}

Figure~\ref{fig:Li+K} is the single-line transmission spectrum of Li\,{\sc i,} showing a flat non-detection result.

Figure~\ref{fig:VO} shows the cross-correlation results of VO. Since the outputs from two VO templates are not consistent with each other, we report them as non-conclusive results.


Figures~\ref{fig:non_detection_ato_ion} and \ref{fig:non_detection_molecular} are the non-detection results of atomic and molecular species from the cross-correlation analyses.

\restartappendixnumbering

\renewcommand{\theHfigure}{A.\arabic{figure}} 
\setcounter{figure}{0}

\begin{figure}
    \centering
    \includegraphics[width=0.99\linewidth]{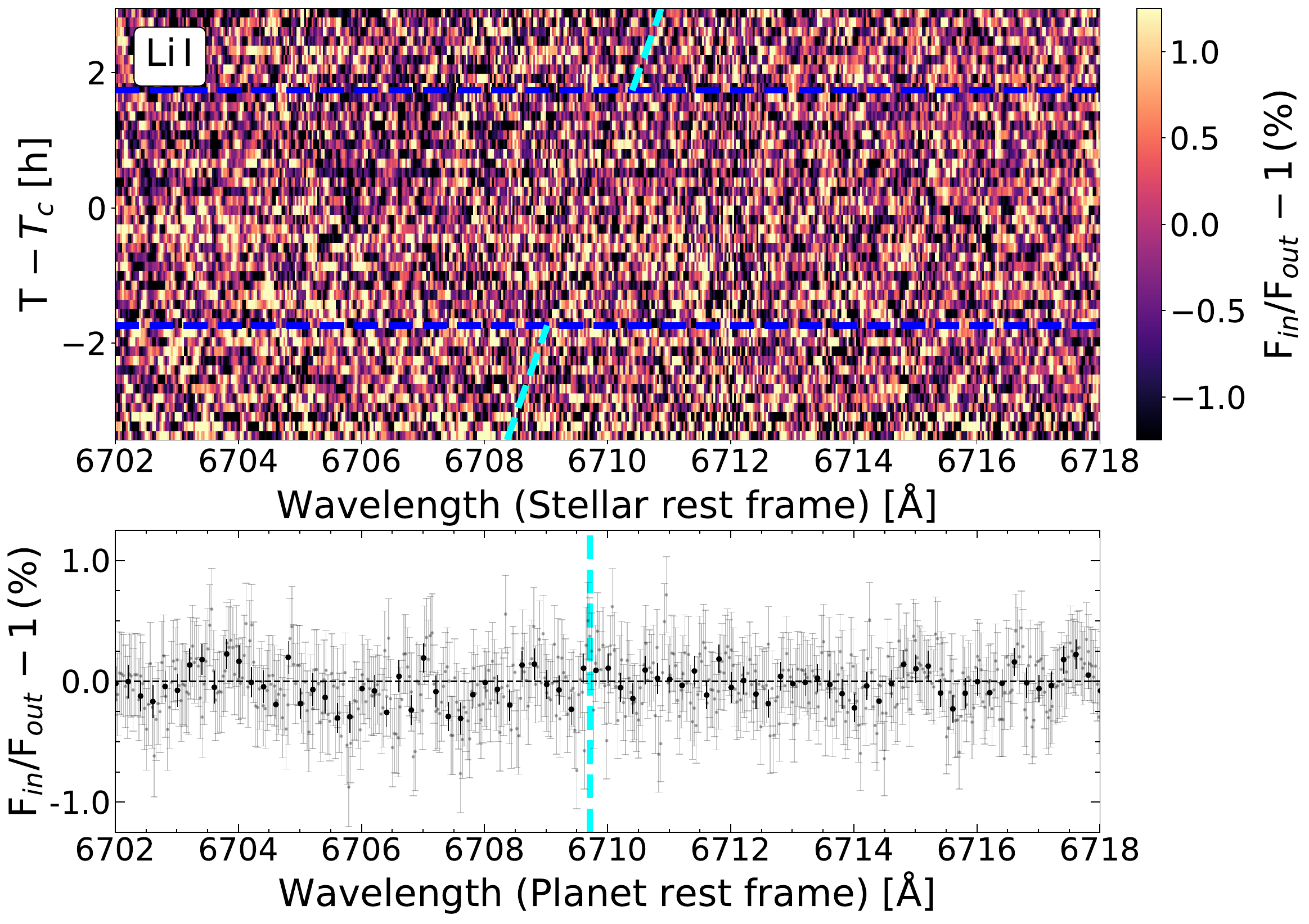}
    \caption{\label{Fig: TS Li K} Same as Figure\,\ref{Fig: TS Halpha Helium} but for Li\,{\sc i} $\lambda$6710\,\AA .
    }
    \label{fig:Li+K}
\end{figure}

\renewcommand{\theHfigure}{A.\arabic{figure}} 
\setcounter{figure}{1}

\begin{figure*}
    \centering
    \includegraphics[width=\textwidth]{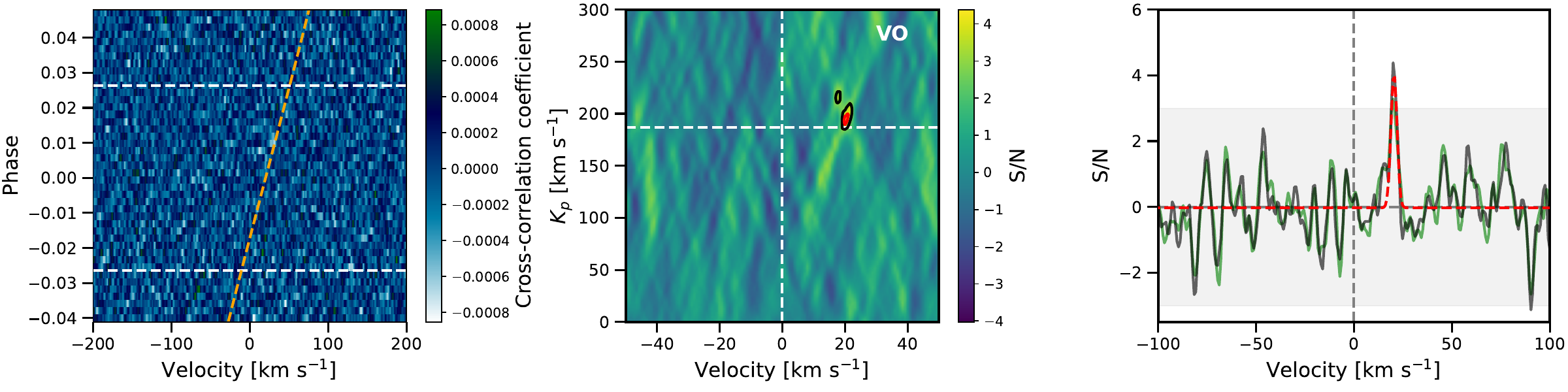}
    \includegraphics[width=\textwidth]{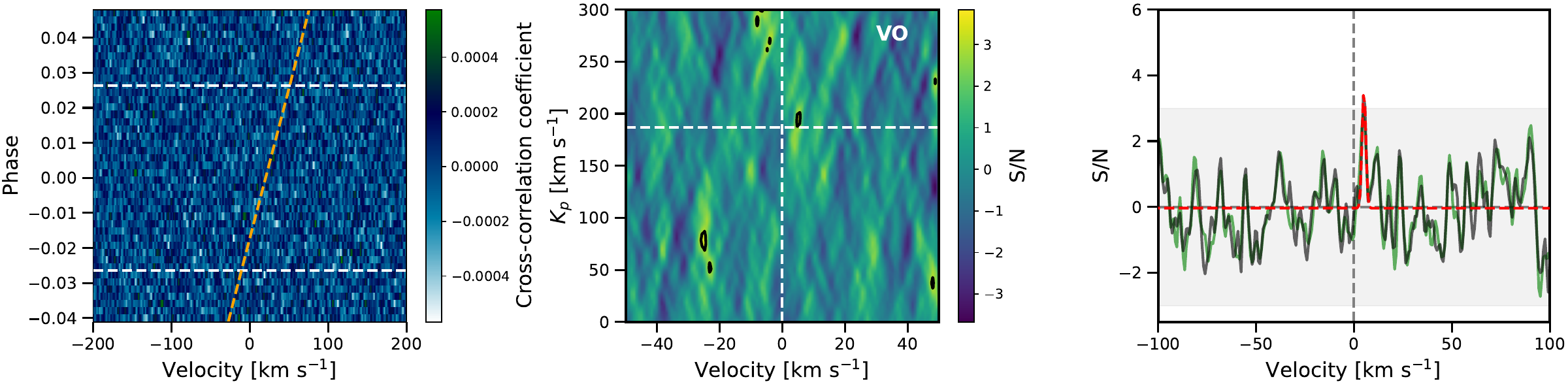}
    \caption{The cross-correlation results of VO using the template from B. Plez \citep[top,][]{Molliere2019} and ExoMol \citep[bottom,][]{Tennyson2016}. The two signals with S/N above 3 are not consistent with each other thus we do not treat them as tentative detections but non-conclusive results.}
    \label{fig:VO}
\end{figure*}



\renewcommand{\theHfigure}{A.\arabic{figure}} 
\setcounter{figure}{2}

\figsetstart
\figsetnum{1}
\figsettitle{Non-detections of atomic species}

\figsetgrpstart
\figsetgrpnum{1.1}
\figsetplot{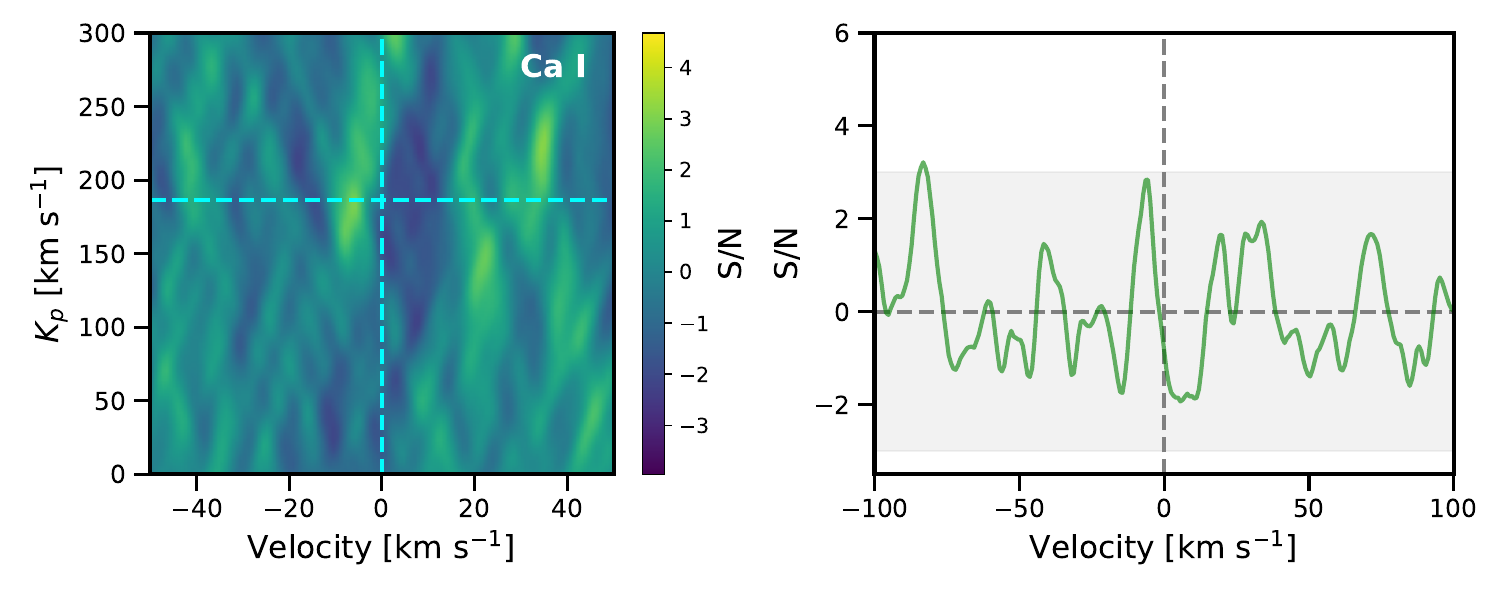}
\figsetgrpnote{The cross-correlation result of Ca\,{\sc i}.}
\figsetgrpend

\figsetgrpstart
\figsetgrpnum{1.2}
\figsetplot{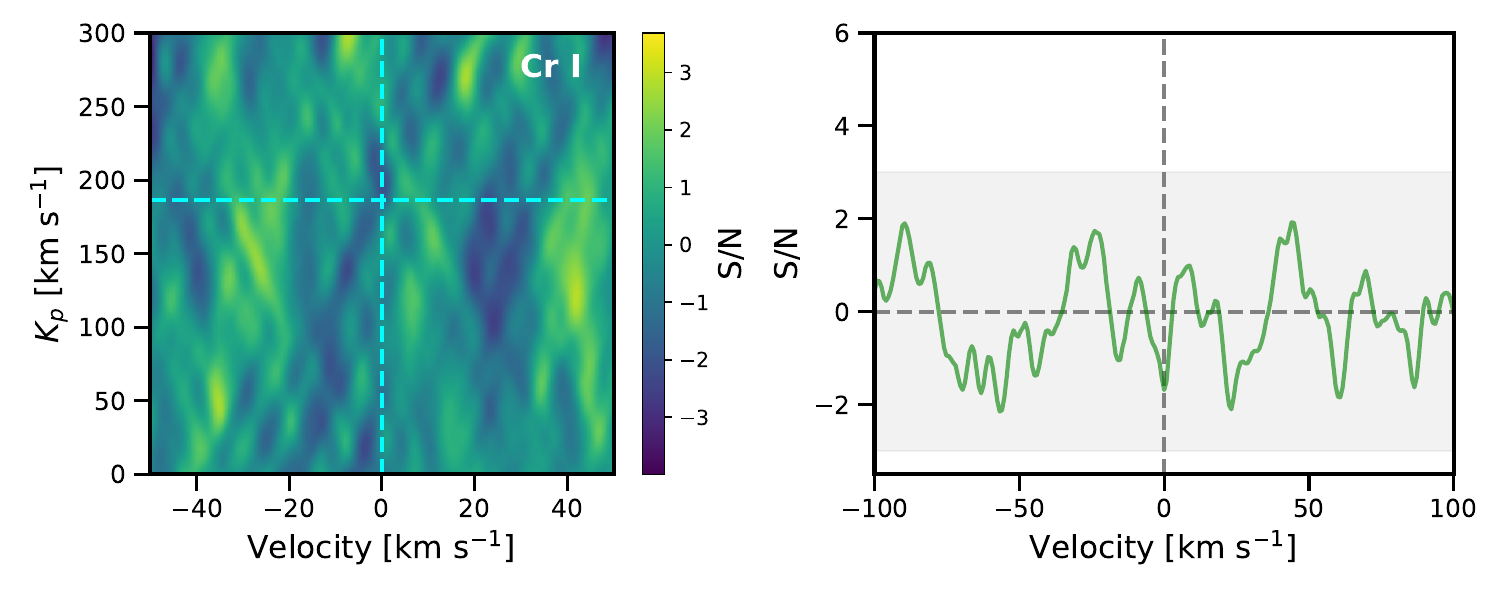}
\figsetgrpnote{The cross-correlation result of Cr\,{\sc i}.}
\figsetgrpend

\figsetgrpstart
\figsetgrpnum{1.3}
\figsetplot{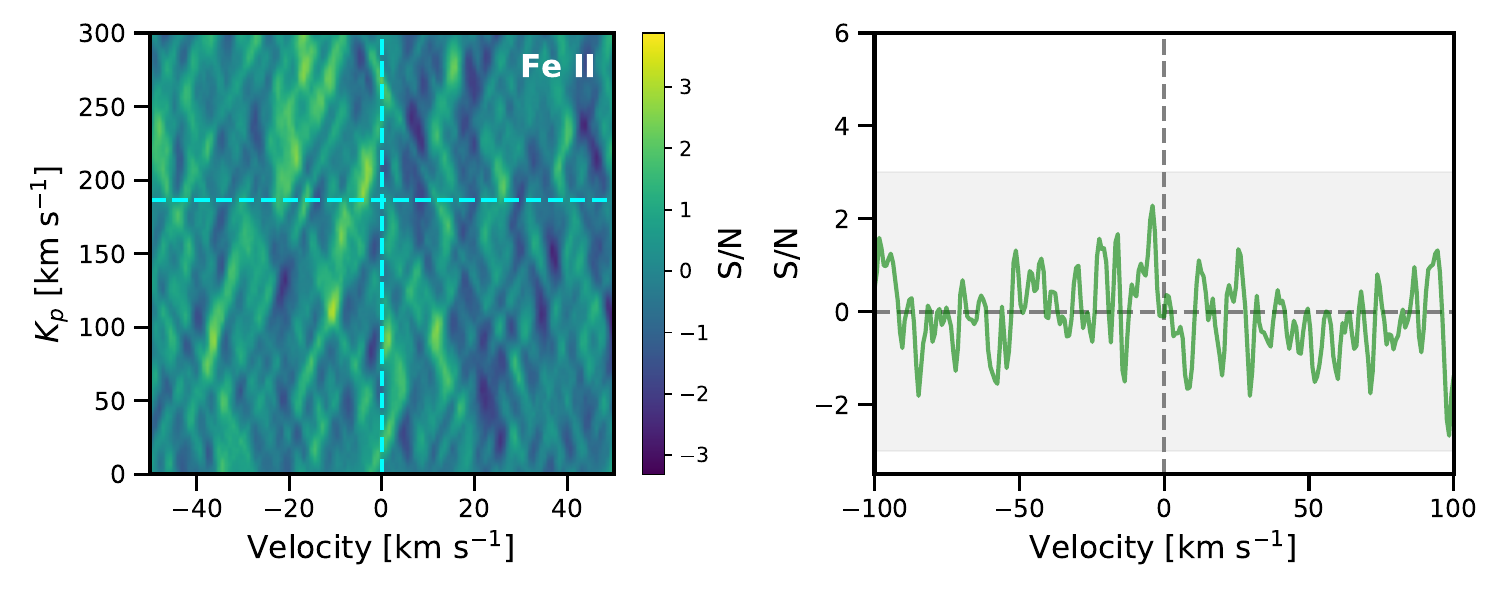}
\figsetgrpnote{The cross-correlation result of Fe\,{\sc ii}.}
\figsetgrpend

\figsetgrpstart
\figsetgrpnum{1.4}
\figsetplot{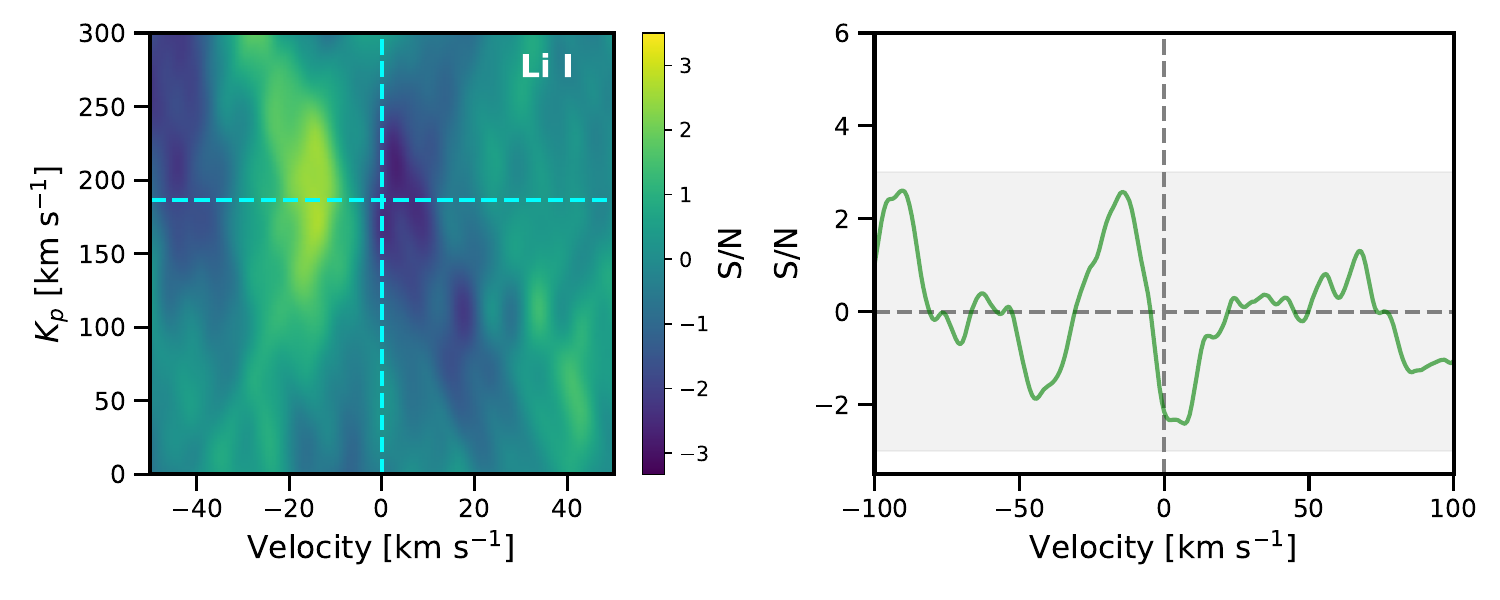}
\figsetgrpnote{The cross-correlation result of Li\,{\sc i}.}
\figsetgrpend

\figsetgrpstart
\figsetgrpnum{1.5}
\figsetplot{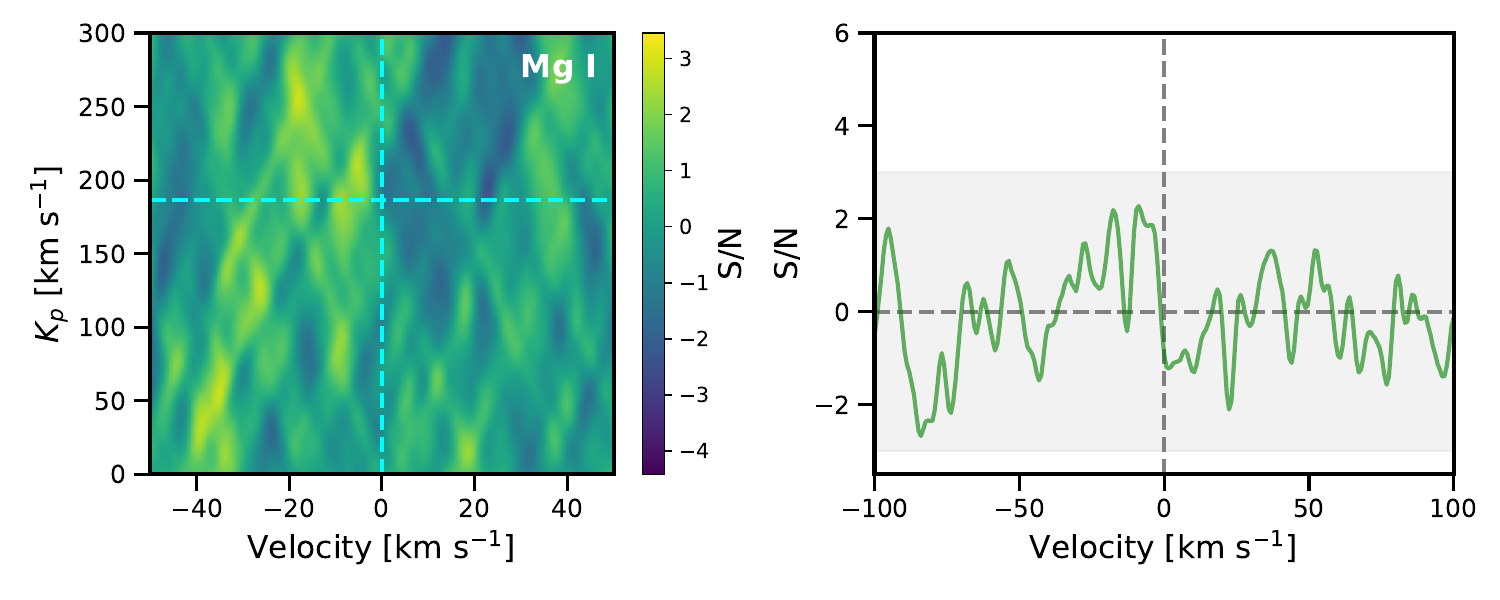}
\figsetgrpnote{The cross-correlation result of Mg\,{\sc i}.}
\figsetgrpend

\figsetgrpstart
\figsetgrpnum{1.6}
\figsetplot{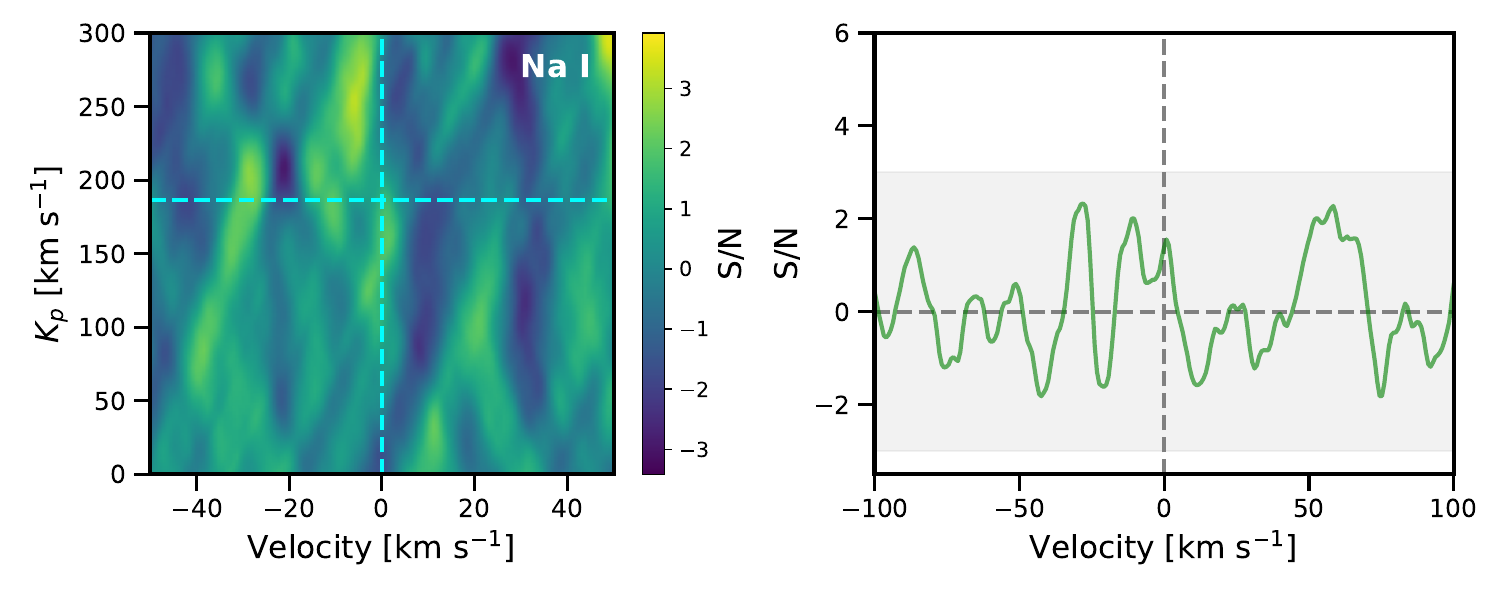}
\figsetgrpnote{The cross-correlation result of Na\,{\sc i}.}
\figsetgrpend

\figsetgrpstart
\figsetgrpnum{1.7}
\figsetplot{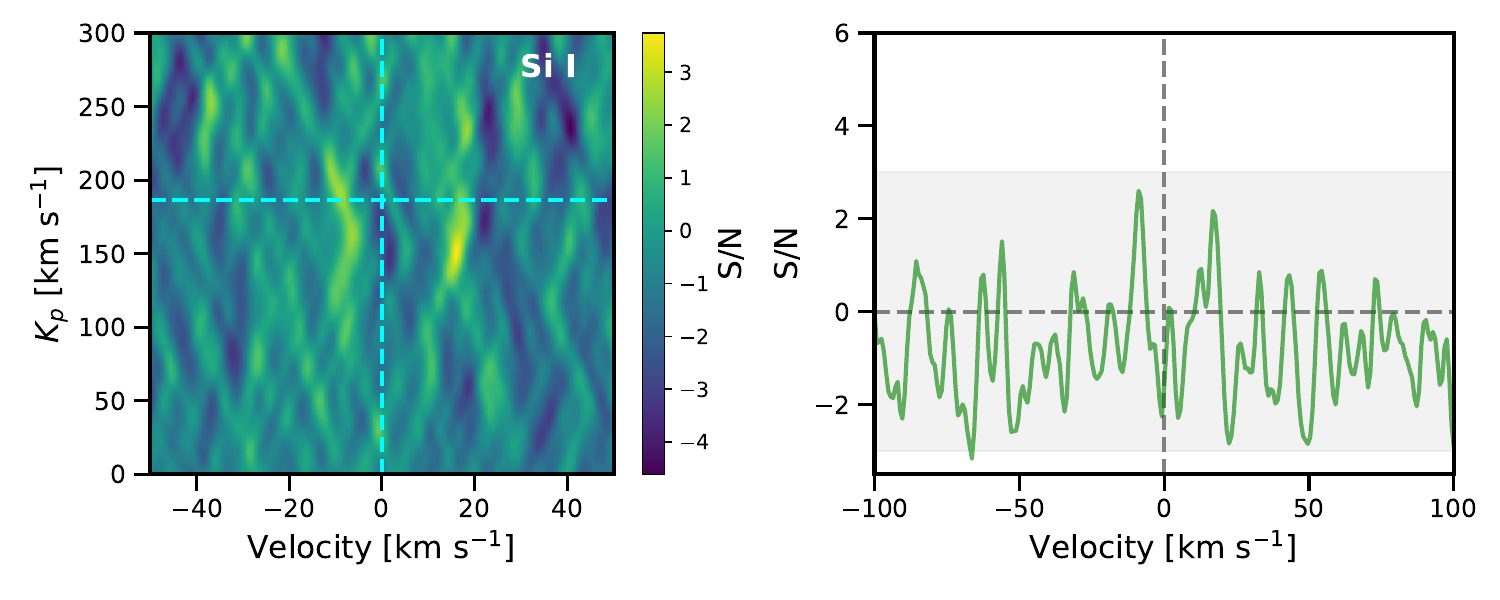}
\figsetgrpnote{The cross-correlation result of Si\,{\sc i}.}
\figsetgrpend

\figsetgrpstart
\figsetgrpnum{1.8}
\figsetplot{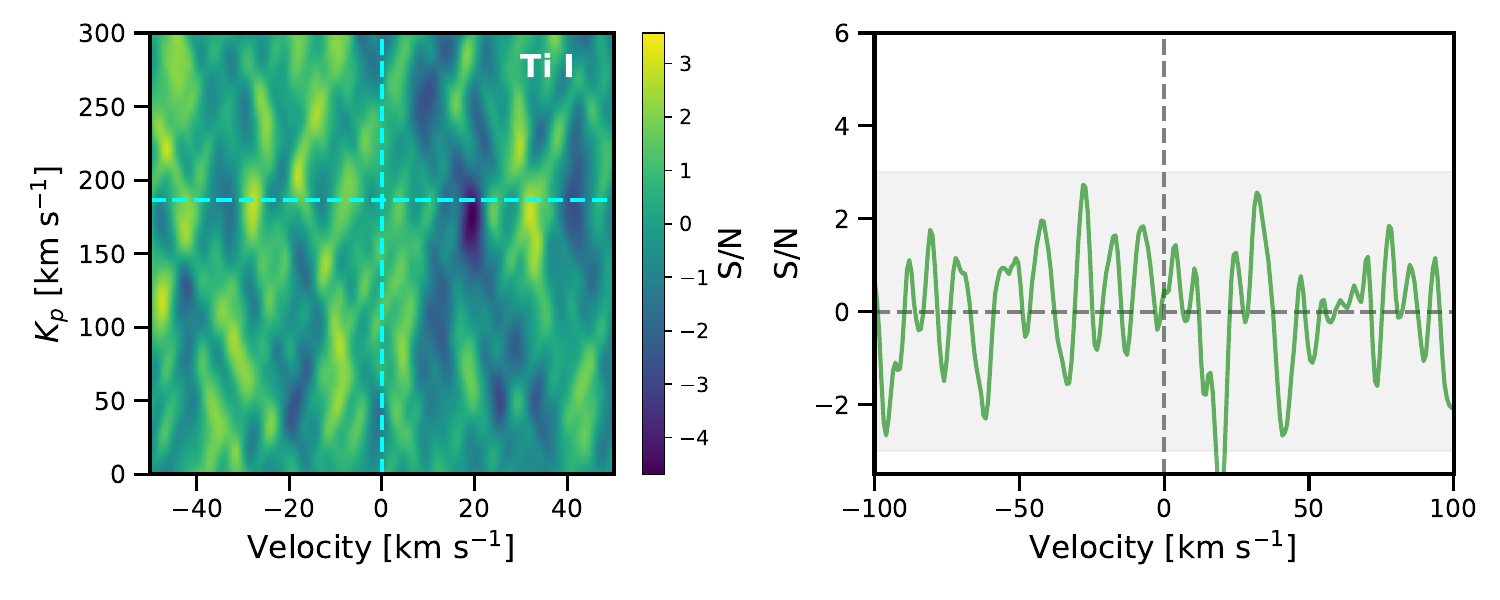}
\figsetgrpnote{The cross-correlation result of Ti\,{\sc i}.}
\figsetgrpend

\figsetgrpstart
\figsetgrpnum{1.9}
\figsetplot{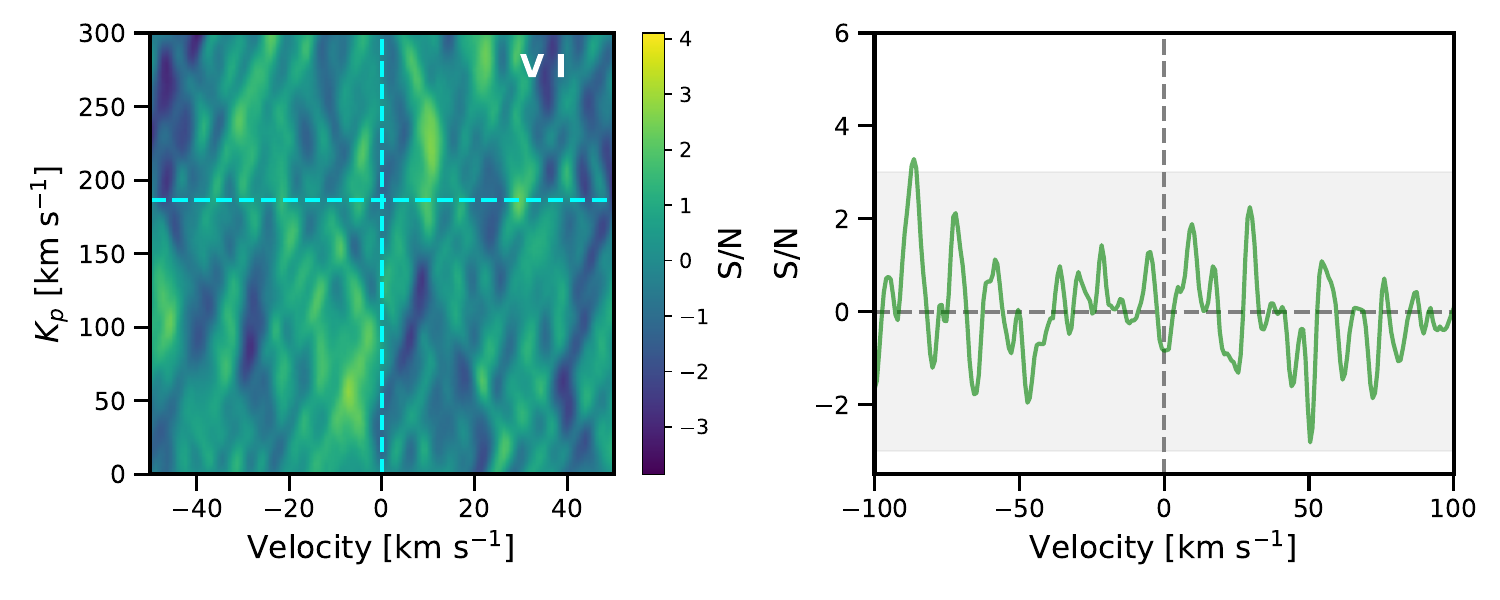}
\figsetgrpnote{The cross-correlation result of V\,{\sc i}.}
\figsetgrpend

\figsetgrpstart
\figsetgrpnum{1.10}
\figsetplot{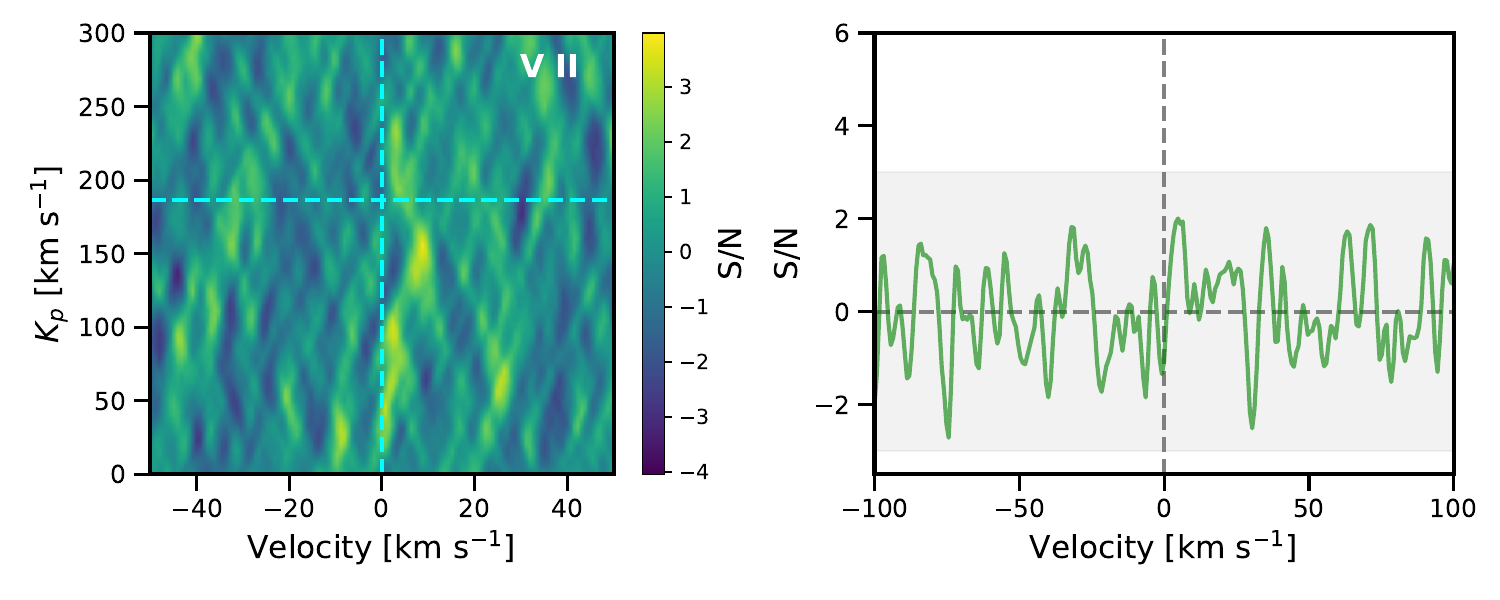}
\figsetgrpnote{The cross-correlation result of V\,{\sc ii}.}
\figsetgrpend

\figsetgrpstart
\figsetgrpnum{1.11}
\figsetplot{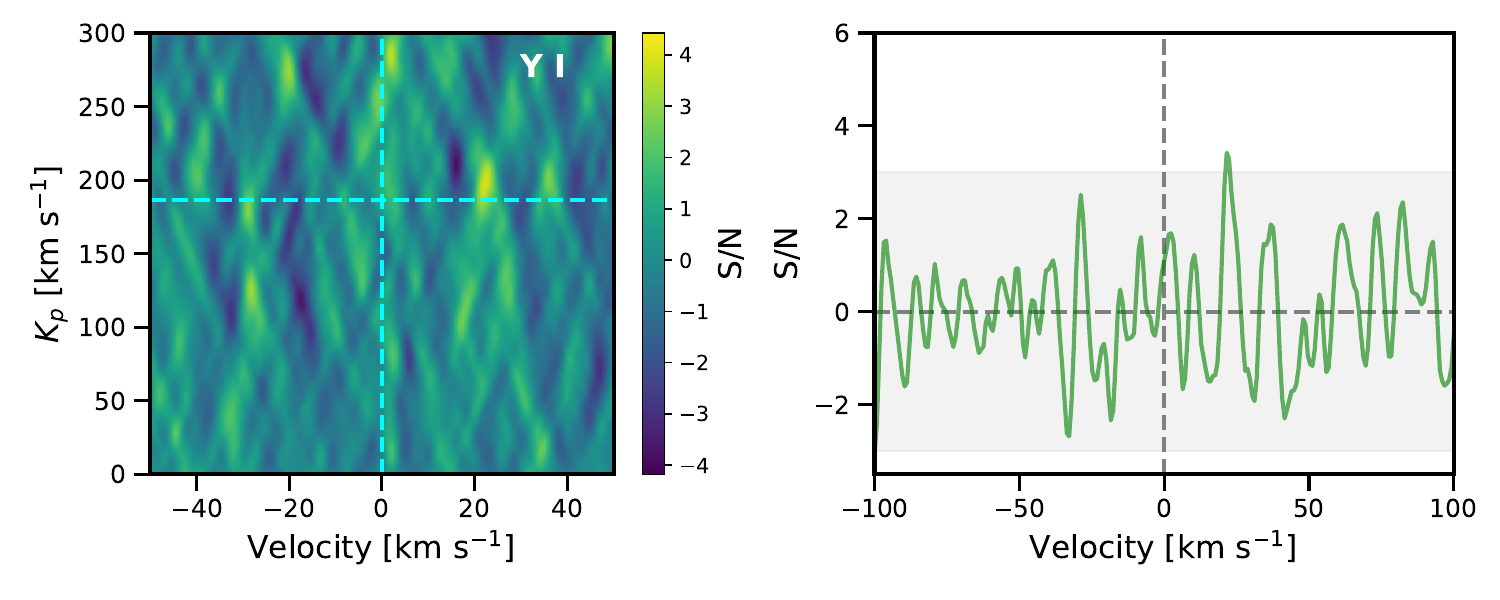}
\figsetgrpnote{The cross-correlation result of Y\,{\sc i}.}
\figsetgrpend

\figsetend

\begin{figure*}
\plotone{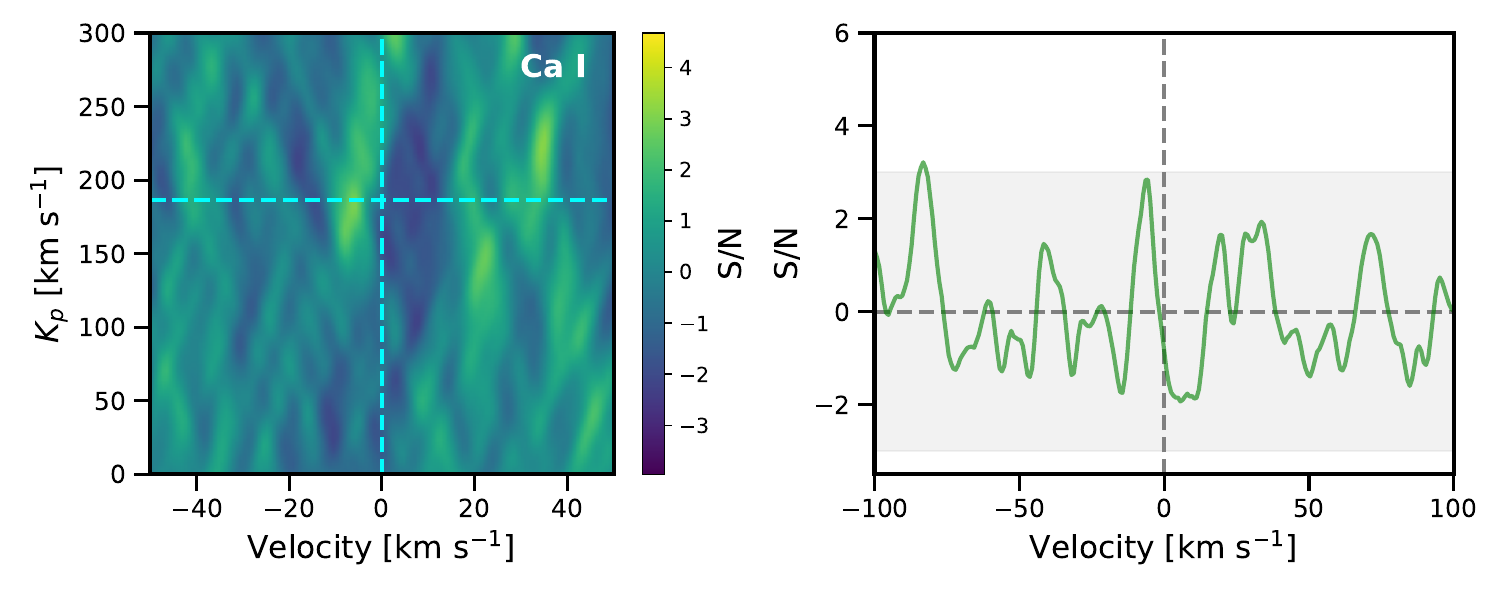}
\caption{Non-detections of 11 atomic species from the cross-correlation analyses. The left panel shows the cross-correlation S/N as a function of $K_p$ and $\Delta v$ while the right panel presents the 1D cross-correlation function at the literature $K_p$. The complete figure set (11 plots) is available in the online journal.}
\label{fig:non_detection_ato_ion}
\end{figure*}

\renewcommand{\theHfigure}{A.\arabic{figure}} 
\setcounter{figure}{3}

\figsetstart
\figsetnum{2}
\figsettitle{Non-detections of molecular species}

\figsetgrpstart
\figsetgrpnum{2.1}
\figsetplot{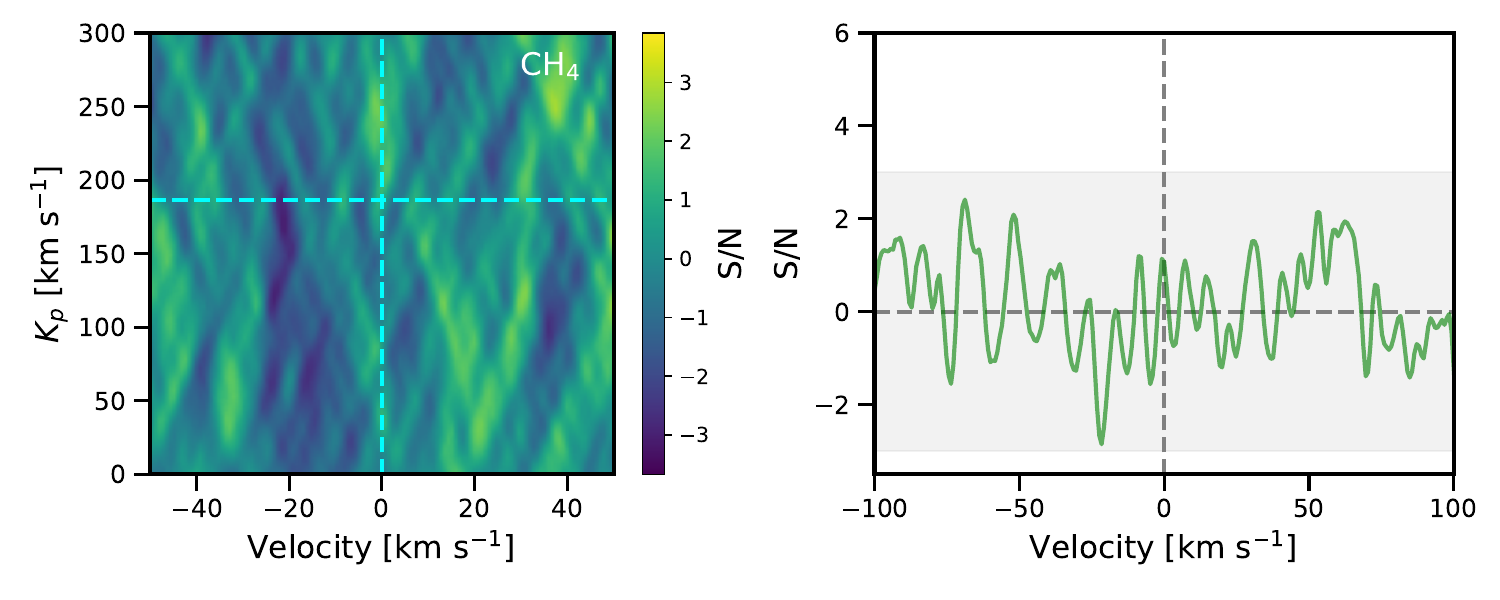}
\figsetgrpnote{The cross-correlation result of $\rm CH_4$.}
\figsetgrpend

\figsetgrpstart
\figsetgrpnum{2.2}
\figsetplot{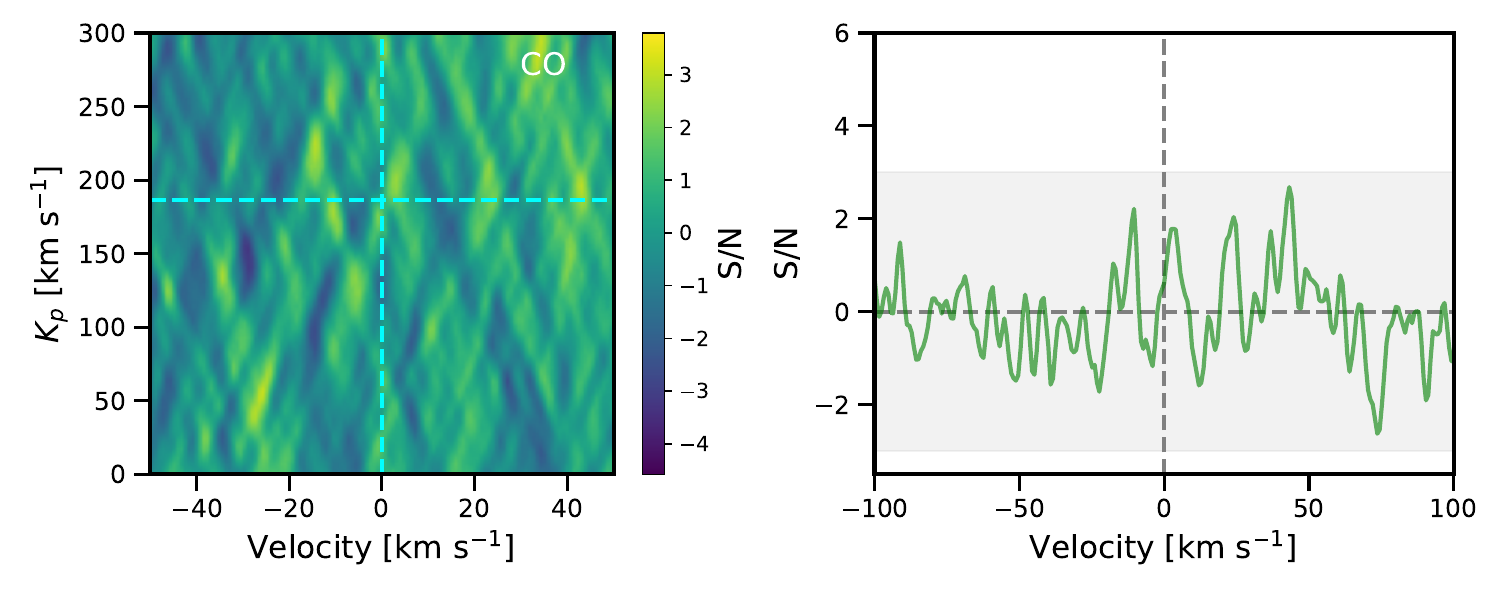}
\figsetgrpnote{The cross-correlation result of $\rm CO$.}
\figsetgrpend

\figsetgrpstart
\figsetgrpnum{2.3}
\figsetplot{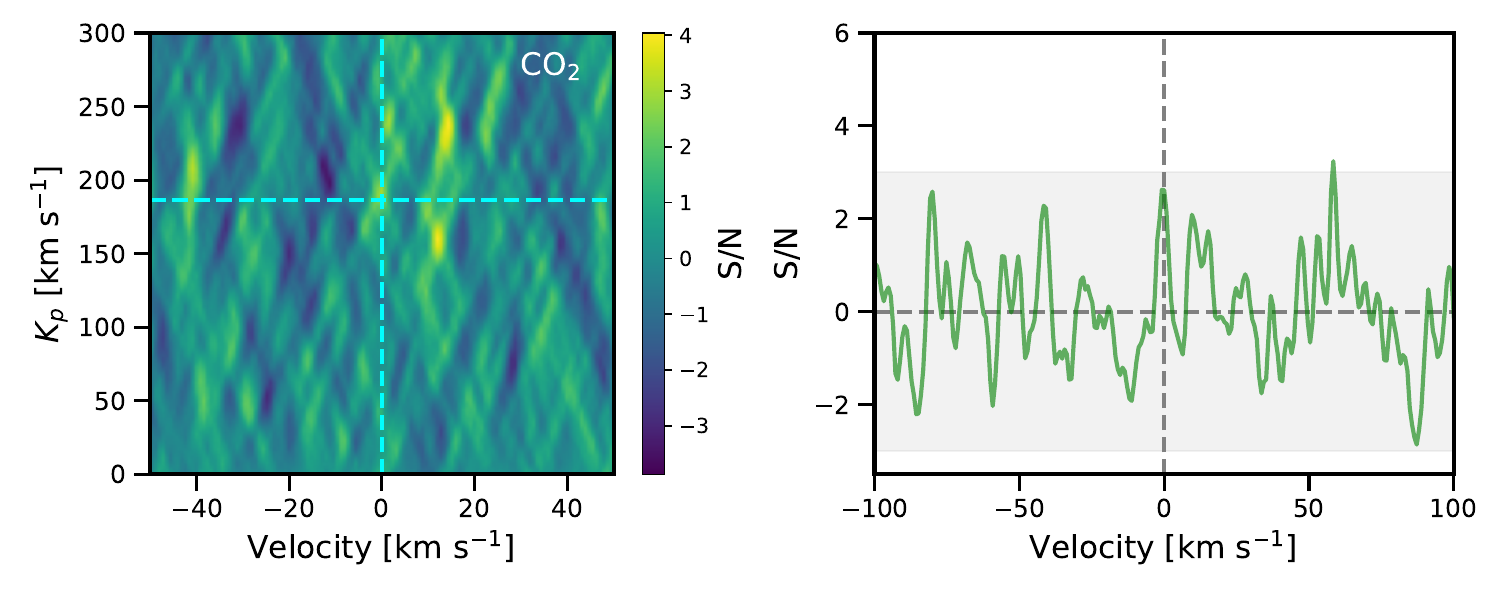}
\figsetgrpnote{The cross-correlation result of $\rm CO_2$.}
\figsetgrpend

\figsetgrpstart
\figsetgrpnum{2.4}
\figsetplot{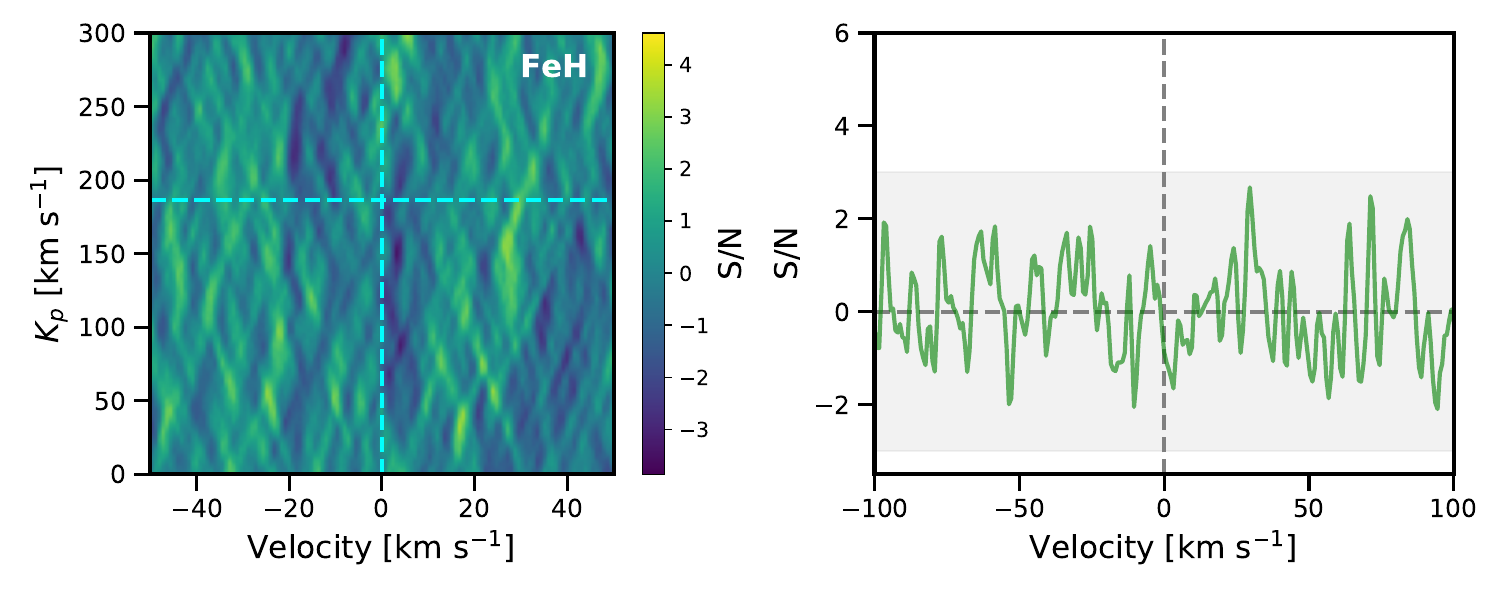}
\figsetgrpnote{The cross-correlation result of $\rm FeH$.}
\figsetgrpend

\figsetgrpstart
\figsetgrpnum{2.5}
\figsetplot{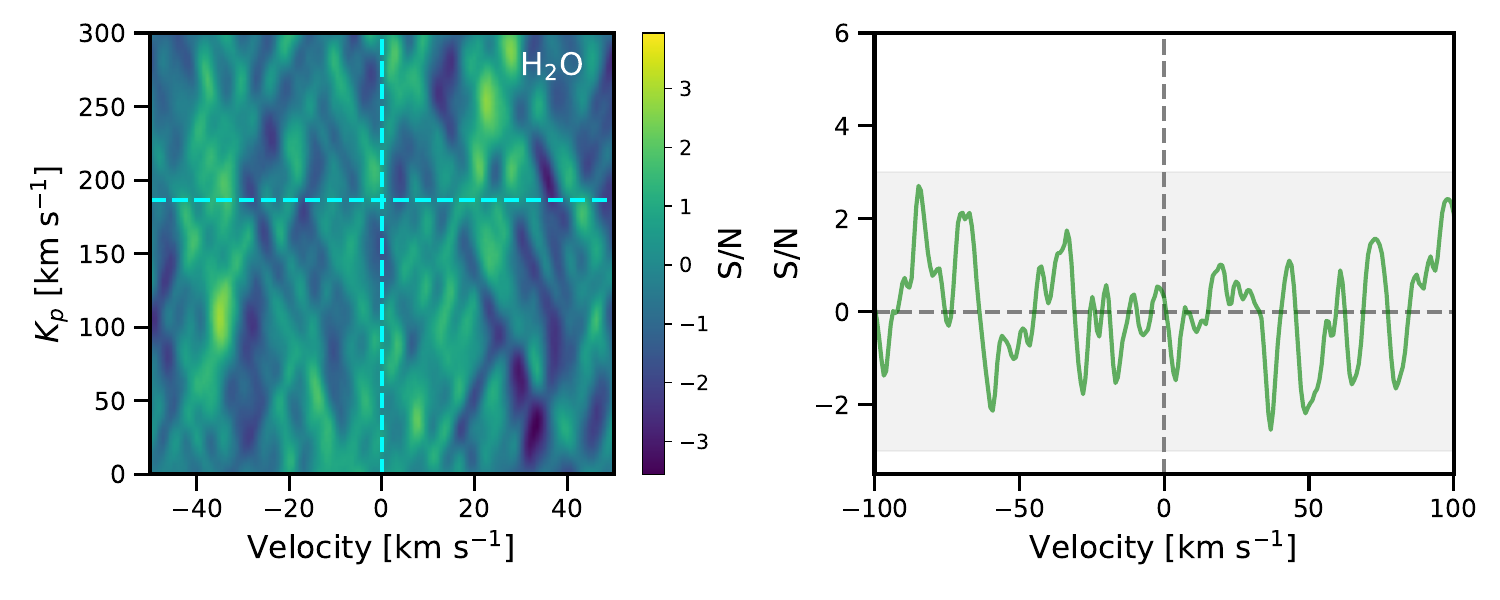}
\figsetgrpnote{The cross-correlation result of $\rm H_2O$.}
\figsetgrpend

\figsetgrpstart
\figsetgrpnum{2.6}
\figsetplot{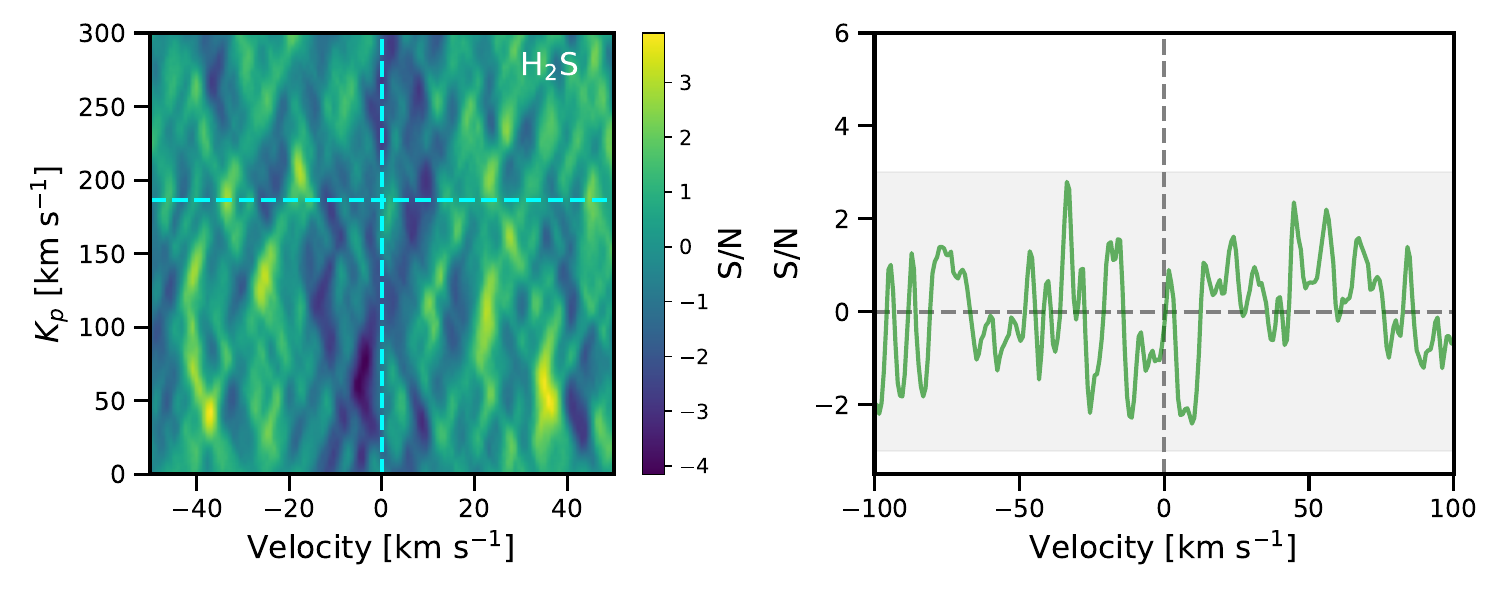}
\figsetgrpnote{The cross-correlation result of $\rm H_2S$.}
\figsetgrpend

\figsetgrpstart
\figsetgrpnum{2.7}
\figsetplot{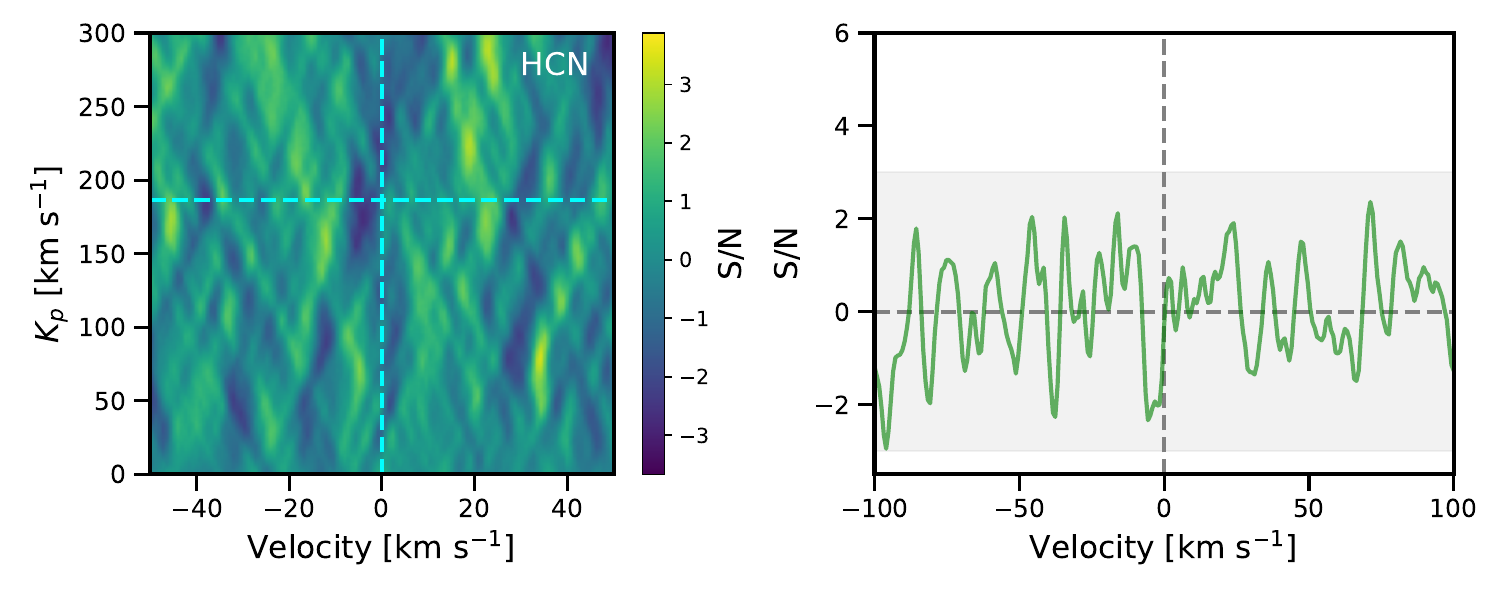}
\figsetgrpnote{The cross-correlation result of $\rm HCN$.}
\figsetgrpend

\figsetgrpstart
\figsetgrpnum{2.8}
\figsetplot{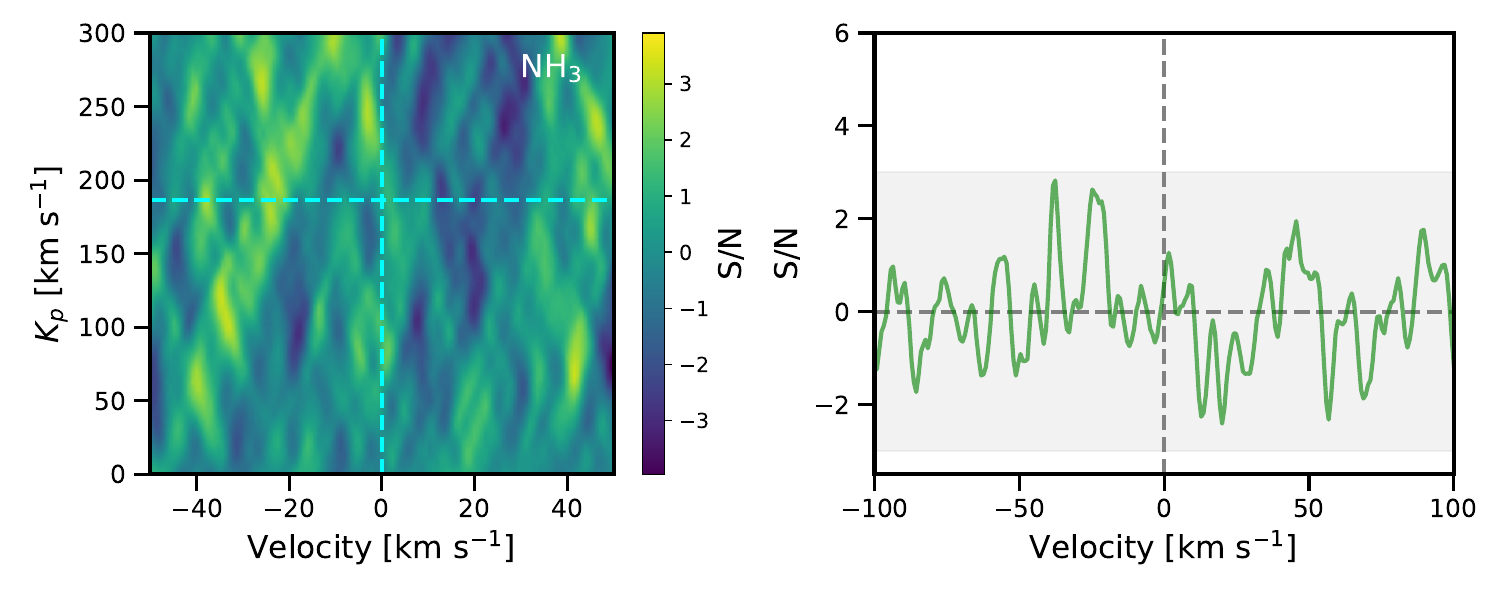}
\figsetgrpnote{The cross-correlation result of $\rm NH_3$.}
\figsetgrpend

\figsetgrpstart
\figsetgrpnum{2.9}
\figsetplot{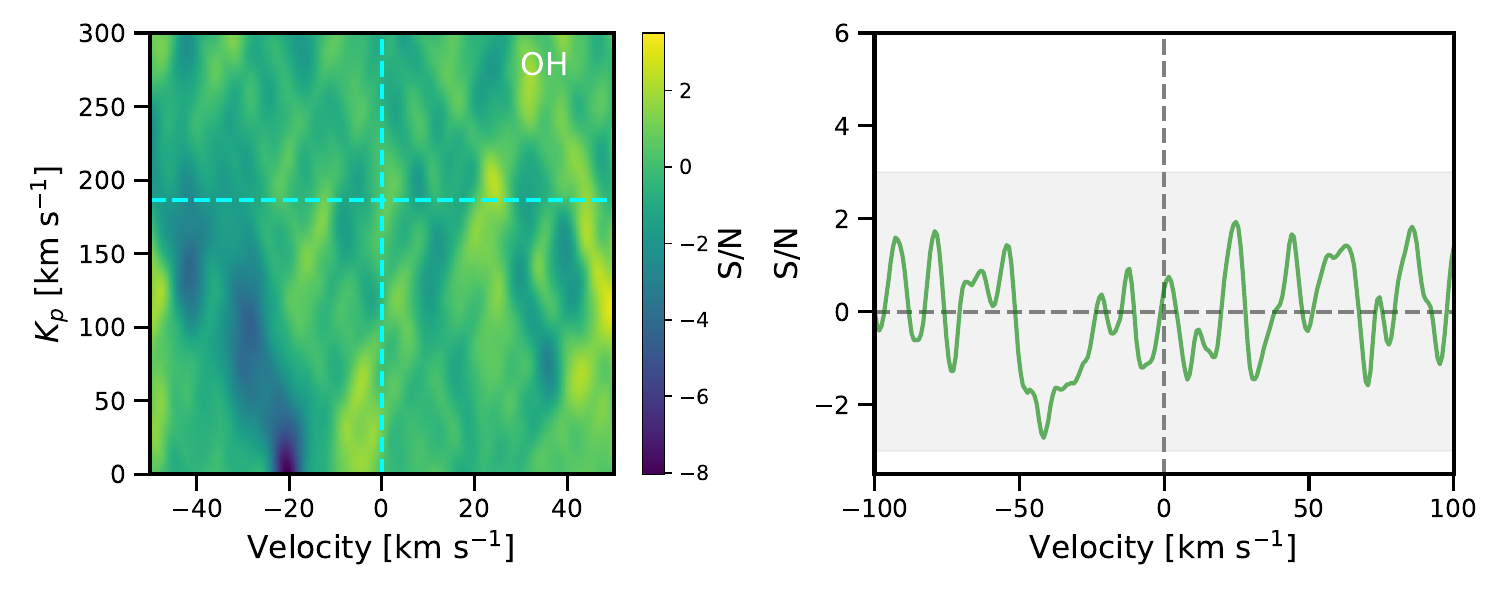}
\figsetgrpnote{The cross-correlation result of $\rm OH$.}
\figsetgrpend

\figsetgrpstart
\figsetgrpnum{2.10}
\figsetplot{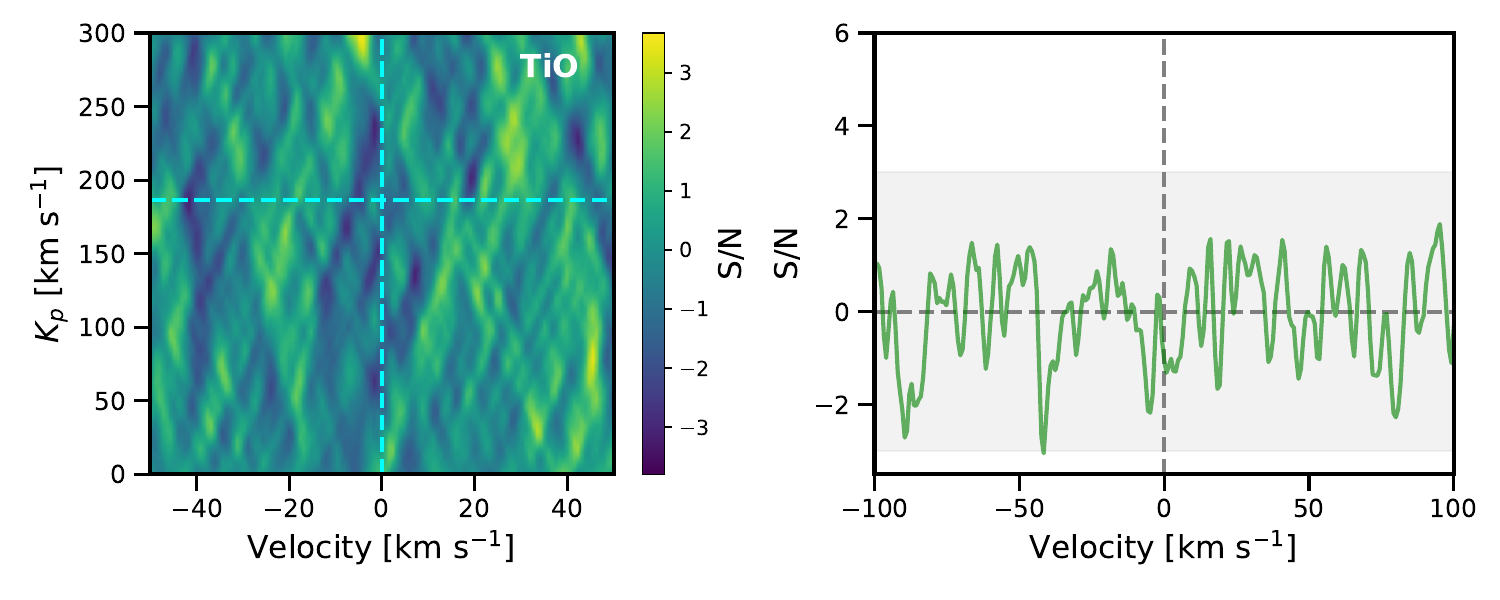}
\figsetgrpnote{The cross-correlation result of $\rm TiO$.}
\figsetgrpend

\figsetend

\begin{figure*}
\plotone{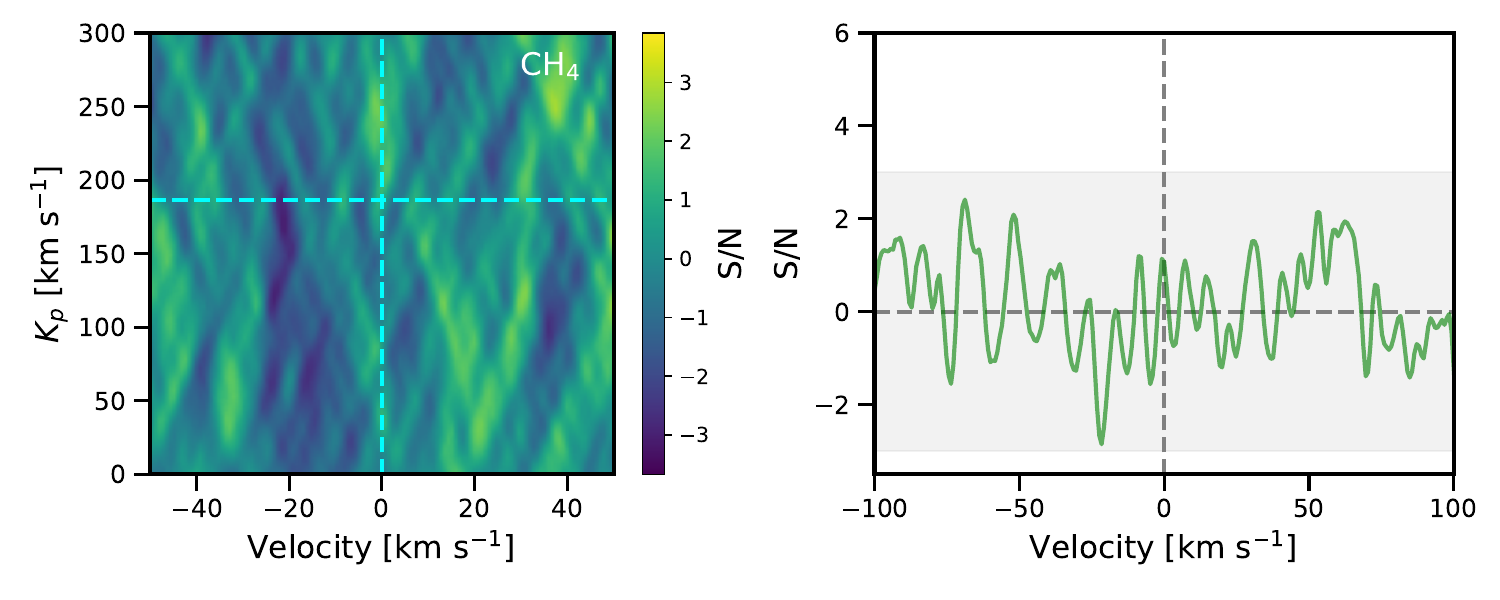}
\caption{Non-detections of 10 molecular species from the cross-correlation analyses. The left panel shows the cross-correlation S/N as a function of $K_p$ and $\Delta v$ while the right panel presents the 1D cross-correlation function at the literature $K_p$. The complete figure set (10 plots) is available in the online journal.}
\label{fig:non_detection_molecular}
\end{figure*}


\bibliography{planet}{}
\bibliographystyle{aasjournal}



\end{document}